\documentclass[a4paper,floatfix,rmp,twocolumn,showkeys,superscriptaddress]{revtex4}             
\usepackage[latin1]{inputenc}
\usepackage{graphicx}
\usepackage{amsmath,amssymb}
\usepackage{extpfeil} 
\usepackage{calrsfs}
\usepackage{yfonts}
\usepackage{comment}
\usepackage{xcolor}
\usepackage{mathrsfs}
\usepackage[colorlinks=true, linkcolor=black, urlcolor=black, citecolor=black]{hyperref}
\usepackage{tikz}
\usetikzlibrary{arrows.meta}
\usepackage{pgfmath}
\usepackage{MnSymbol}
\usepackage{tikz}
\usepackage{enumitem}
\usetikzlibrary{arrows.meta,positioning}
\usetikzlibrary{decorations.pathreplacing}
\usetikzlibrary{arrows.meta,positioning,calc}

\newif\ifarxiv
\arxivtrue

\usepackage{color}
\usepackage{newtxtext}

\newcommand\smrefnocite[1]{SM-\ref*{#1}}

\ifarxiv
 \let\savedaddcontentsline\addcontentsline 
\global\long\def\addcontentsline#1#2#3{}%

\else

\usepackage{xr}

\newcommand*{\addFileDependency}[1]{%
\typeout{(#1)}%
\@addtofilelist{#1}
\IfFileExists{#1}{}{\typeout{No file #1.}}
}
\newcommand*{\myexternaldocument}[1]{%
\externaldocument{#1}%
\addFileDependency{#1.tex}%
\addFileDependency{#1.aux}%
}

\myexternaldocument{sm}

\fi
\usepackage[most]{tcolorbox}

\newcounter{mybox}
\newtcolorbox[use counter=mybox]{mytextbox}[1]{%
  colback=gray!10,      
  colframe=black,       
  colbacktitle=gray!0,  
  coltitle=black,       
  fonttitle=\bfseries,  
  title=BOX~\themybox:~#1,
  boxrule=0.5pt,
  arc=0.5mm,
  left=1mm,
  right=1mm,
  top=1mm,
  bottom=1mm
}

\allowdisplaybreaks
\makeatother

\begin{document}

\makeatletter
\renewcommand\@biblabel[1]{[#1]}
\makeatother
\setcitestyle{numbers,square}

\makeatletter
\global\long\def\mysection#1{%
\emph{#1}.--- %
\@ifnextchar\par{\@gobble}{}}%
\makeatother


\global\long\def\bS{\boldsymbol{S}}
\global\long\def\bh{\boldsymbol{h}}
\global\long\def\bQ{\boldsymbol{q}}
\global\long\def\bm{\boldsymbol{m}}
\global\long\def\bq{\boldsymbol{q}}
\global\long\def\1{\boldsymbol{1}}
\global\long\def\0{\boldsymbol{0}}
\global\long\def\bXi{\boldsymbol{\Xi}}
\global\long\def\bxit{\boldsymbol{\xi}^{(\textrm{T})}}
\global\long\def\bxip{\boldsymbol{\xi}^{(\textrm{G})}}
\global\long\def\bxis{\boldsymbol{\xi}^{(\textrm{S})}}
\global\long\def\xit{\xi^{(\textrm{T})}}
\global\long\def\xis{\xi^{(\textrm{S})}}
\global\long\def\hor{\Delta}
\global\long\def\train{\mathcal{T}}
\global\long\def\test{\tilde{\mathcal{T}}}
\global\long\def\tphys{t^{\textrm{phys}}}
\global\long\def\bz{\boldsymbol{z}}
\global\long\def\bs{\boldsymbol{s}}
\global\long\def\bximu{\boldsymbol{\xi}^{(\mu)}}
\global\long\def\bximuT{{\boldsymbol{\xi}^{(\mu)}}^{\intercal}}
\global\long\def\bxinu{\boldsymbol{\xi}^{(\nu)}}
\global\long\def\brho{\boldsymbol{\rho}}
\global\def\mix{\Lambda}
\global\long\def\F{\mathfrak{F}}
\global\long\def\Reals{\mathbb{R}}
\global\long\def\mP{m_{\textrm{G}}}
\global\long\def\mS{m_{\textrm{S}}}
\global\long\def\mT{m_{\textrm{T}}}


\title{Nonreciprocal Hopfield nets for angiogenesis}

\providecommand{\UPC}{Department of Mathematics, Universitat Polit\`ecnica de Catalunya, 08034 Barcelona, Spain}
\providecommand{\MSUCOMP}{Department of Computational Mathematics, Science and Engineering, Michigan State University, East Lansing, Michigan 48824, USA}
\providecommand{\TCD}{Department of Genetics, Trinity College Dublin, D02 VF25 Dublin, Ireland}
\providecommand{\MSUPHYS}{Department of Physics and Astronomy, Michigan State University, East Lansing, Michigan 48824, USA}
\providecommand{\UMA}{Department of Molecular Biology and Biochemistry, University of M\'alaga, Andaluc\'ia Tech, E-29071 M\'alaga, Spain}

\author{Jordi Pi\~nero}
\email{jordi.pinero@upc.edu}
\affiliation{\TCD}
\affiliation{\UPC}
\author{Krishna Chaitanya Ginne}
\affiliation{\MSUPHYS}
\author{Antonio Alves Ducci}
\affiliation{\MSUPHYS}
\author{Ra\'ul Monta\~nez}
\affiliation{\UMA}
\author{Carlo Piermarocchi}
\email{piermaro@msu.edu}
\affiliation{\MSUPHYS}
\affiliation{\MSUCOMP}

\date{\today}

\begin{abstract}

During sprouting angiogenesis, endothelial cells dynamically reorganize
among three relevant phenotypes (glycolytic, stalk and tip) to form new blood vessels, with cells transitioning from one phenotype to another. These cell-state transitions may involve intrinsic nonequilibrium dynamics that
go beyond relaxation of the conventional Waddington landscape. We
investigate this possibility using a nonreciprocal Hopfield network inferred
from single-cell RNA sequencing and RNA velocity data for 1896
endothelial cells on a panel of 1873 genes. We encode the three experimentally identified
phenotypes as correlated memory patterns and introduce an asymmetric
mixing matrix describing their effective interactions. A dynamical mean-
field reduction yields a three-dimensional system for the overlaps with
these phenotypic memories, whose parameters are fitted to RNA-
velocity-derived trajectories. The nonreciprocal model outperforms its
reciprocal counterpart on held-out testing data. Remarkably, its long-time
dynamics exhibit either stationary attractors or sustained cycles
connecting phenotypic states, depending on the inference timescale.
These results suggest that angiogenic cell fate decisions may involve
directed, non-relaxational dynamics that cannot be captured by a
conventional energy landscape. This approach may also provide a useful framework for investigating how perturbations constrain cellular differentiation.

\end{abstract}

\keywords{Cell Regulation, Angiogenesis, Nonreciprocal Networks, Waddington Landscape}

\maketitle

\section{Introduction}
\label{sec:intro}

In multicellular organisms, transitions between cellular phenotypic identities are governed by complex transcriptional programs~\cite{davidson2010regulatory}. These systems dynamically control the evolution of the cellular state~\cite{britten1969gene} in concert with extrinsic factors such as cell-cell signaling, positional information, and morphogenetic gradients~\cite{webster1996form,wolpert2015principles}. Traditionally, cell fate determination has been conceptualized through Waddington's landscape~\cite{waddington2014strategy}, wherein cell identities are depicted as basins of attraction on a potential surface~\cite{huang2005cell, huang2012molecular}. This picture captures the robustness of developmental trajectories and phenotypic outcomes through homeorhesis and canalization~\cite{waddington2014strategy}. Advances in single-cell profiling have also revealed extensive transcriptional heterogeneity within and between phenotypic states~\cite{svensson2018exponential,tanay2017scaling}, emphasizing the pervasive biological variability described as \emph{poikilosis}~\cite{vihinen2020poikilosis}. However, an account of phenotypic robustness does not establish purely relaxational dynamics capable of connecting cellular states in a driven, directed manner. In a deterministic view, descent on a fixed potential excludes autonomous sustained cycles, while in a stochastic set-up, equilibrium fluctuations permit switching but only without the presence of persistent probability currents. The central question is therefore whether relaxation and fluctuations sufficiently describe phenotypic plasticity, or whether effective models must also accommodate nonrelaxational driving~\cite{zhu2024uncovering}.

In recent geometric approaches to this problem~\cite{rand2021geometry,cislo2025reconstructing}, authors fit dynamical landscapes to single-cell data through a potential function coupled to a nontrivial underlying metric by drawing on Morse--Smale theory. On the other hand, in~\cite{zhu2024uncovering}, Wang and colleagues reconstruct landscape and nonequilibrium flux contributions from vector fields inferred using RNA velocity. These approaches depart from the assumption that a potential alone determines the dynamics of cell fate by complementing the landscape framework. 
In general, a landscape description coupled with nonzero probability currents requires additional dynamical information that exceeds the modeling capacities of the Waddington framework.

\begin{figure}
    \centering
    \includegraphics[width=\linewidth]{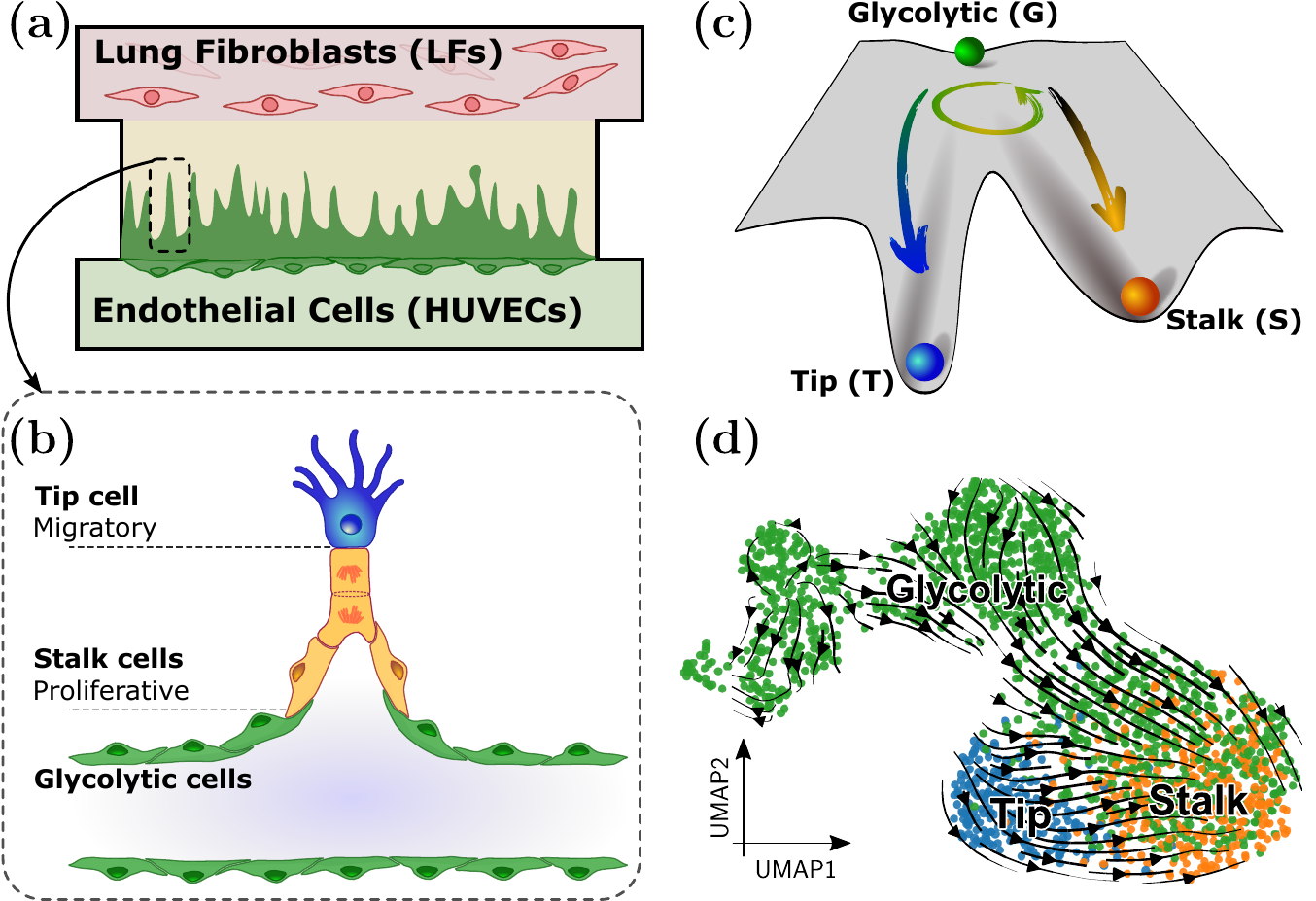}
    \caption{{\bf (a) Angiogenesis experiments.} Scheme of set-up employed by Lee {\em et al.} in~\cite{lee2024angiogenesis}. A three stage chip is used with lung fibroblasts (LFs) on the top and endothelial cells (ECs) at the bottom. Sprouting occurs in the middle stage as a result of the biomolecular signaling emmitted from the LFs. {\bf (b) EC phenotpyes in the EndMT.} Newly formed sprouts involve 
    three phenotypic states: {\em {glycolytic}} (G) cells, which are queiscent and stabilize the main vessel; {\em stalk} (S) cells, which are proliferative and conform the walls of the newly form vessel; and {\em tip} (T) cells, which behave as mesenchymal (mobile) cells, grow filopodia and guide sprout formation. {\bf (c) Waddington landscape \& nonconservative currents.} The landscape approach assumes only relaxational dynamics (i.e., descend on a potential surface), but does not account for nonrelaxational modes such as cyclic forces.  {\bf (d) RNA velocity projection.} From scRNA-seq, recent methods allow to estimate the direction and speed in expression space. Here, we color cells according to their phenotype P (green), S (orange) and T (blue). This depiction is a UMAP projection with the RNA velocity field overlayed on top.}
    \label{fig:FIG1}
\end{figure}

Associative-memory networks provide a statistical framework connecting expression patterns to collective cellular states, for example, in cellular cycles~\cite{szedlak2017cell}. A central model for associative networks is based on the Hopfield model, which encodes data patterns in the interactions among many degrees of freedom (nodes) of a network~\cite{hopfield1982neural}. Hopfield nets have been applied to cellular identity and reprogramming, including asymmetric developmental couplings~\cite{szedlak2014control,lang2014epigenetic}. 
Nonreciprocal (or asymmetric) Hopfield couplings allow us to incorporate differences in the directed influences between pairs of components. This directionality is biologically plausible: regulation of one gene by another need not imply an equivalent reverse influence~\cite{davidson2010regulatory}. Nonreciprocal Hopfield nets are considered an extension of Hopfield's original (symmetric) model, and one of their  distinctive features is their ability to support persistent cycles~\cite{kleinfeld1986sequential,sompolinsky1986temporal,amit1988neural}. This feature is particularly relevant to biological systems, where stable developmental and regulatory programs often involve a directed progression through a sequence of cellular states, consistent with the concept of homeorhesis, rather than convergence to a fixed point. Nonreciprocal networks can intrinsically encode such dynamical programs, providing a modeling framework that does not rely exclusively on stochastic fluctuations or external signals, although both remain important components of biological regulation. In this work, we will compare the validity and performance of reciprocal and nonreciprocal models in the context of phenotype dynamics. 

Sprouting angiogenesis provides a suitable setting for studying this phenotypic plasticity. 
In angiogenesis, endothelial cells reorganize their phenotypic profiles to produce new vessels (sprouts) that emerge from the existing vascular tissue. As such, angiogenesis constitutes a relevant example of tissue remodeling and repair, with major implications in the context of tumor vascularization~\cite{ferrara2003biology,carmeliet2003angiogenesis}. In angiogenesis, endothelial cells acquire phenotypic features that facilitate the formation of sprouts, with a specialized migratory tip phenotype (known as {\em tip} cell) that drags the newly formed vessel from its tip, and a proliferative trailing phenotype (known as {\em stalk} cell) that fills the void created by the tip (see Fig.~\ref{fig:FIG1}a-b). These phenotypes are dynamically regulated during sprout formation~\cite{jakobsson2010endothelial, kang2024spatial}. Endothelial plasticity also encompasses the endothelial-to-mesenchymal transition (EndMT), in which endothelial and mesenchymal features coexist~\cite{fang2021regulation,li2025regulation}. The EndMT is usually characterized within a broader range of developmental and disease-associated transitions~\cite{kovacic2019endothelial,qian2025broadening}. Here, we focus on the endothelial expression states resolved in an experimentally controlled angiogenesis assay.

In this paper, we analyze scRNA-seq data from the angiogenesis-on-a-chip experiments of Lee {\em et al.}~\cite{lee2024angiogenesis}, in which human umbilical vein endothelial cells form sprouts in response to signals from co-cultured lung fibroblasts (Fig.~\ref{fig:FIG1}a-b). Our processed dataset contains $1896$ endothelial cells that are phenotypically annotated according to glycolytic, stalk, or tip phenotypes using a common panel of $1873$ genes. RNA velocity supplements each expression profile with an estimate of the local rate of change of spliced-transcript abundance~\cite{la2018rna,bergen2020generalizing}. These estimates provide dynamical constraints from a snapshot ensemble, while subsequent expression changes are inferred from the static data.

Using these data, we infer a nonreciprocal Hopfield model whose empirical memory patterns represent the three phenotypes. A dynamical mean-field reduction describes the evolution through three overlap parameters, whose evolution is governed by a $3\times3$ mixing matrix. We estimate these nine entry values from finite-time targets, which we construct by data extrapolation using RNA velocity.  
We compare our fit against an independently fitted reciprocal model and show improved agreement with held-out data in favor of the nonreciprocal extension. The inferred dynamics also exhibit persistent phenotypic cycles, which appear within a range of biologically plausible parameters. This result suggests a potential regulatory mechanism in which phenotypic indeterminacy is utilized by the system to increase plasticity and phenotypic availability in tissue remodeling, in our case for vascularization.

Our results provide a set of couplings that effectively describe the gene-to-gene interactions under the conditions of the experiment. Our aim is to produce a tractable statistical description that connects high-dimensional expression data with a predictive framework for collective phenotype dynamics. With our framework, we provide better tools to approach mechanistic models in which encoded gene-to-gene relations map onto concrete molecular pathways, as emphasized in recent discussions of how to approach inference in gene regulatory networks~\cite{maizels2026gene}. 
As we briefly discuss in the final section of this paper, our framework can also be useful for investigating gene-network responses to perturbations.

\section{DATA PROCESSING}
\label{sec:data}

In this work, we utilize single-cell RNA sequencing (scRNA-seq) data from an experimentally controlled sprouting angiogenesis process to capture the phenotypic heterogeneity of the endothelial cell (EC) population and probe its dynamics through a nonrelaxational modeling framework. Our analysis builds on the dataset generated by Lee {\em et al.} (2024)~\cite{lee2024angiogenesis}, who cultured human umbilical vein endothelial cells (HUVECs) in a 3D microfluidic device capable of hosting angiogenic expansion across an extracellular matrix layer. HUVECs were maintained 
alongside human lung fibroblasts (LFs) in the opposing chamber to provide 
angiogenic signaling that induces the formation of endothelial sprouts. Five days after loading the chip, cells were harvested and transcriptionally profiled. While Lee {\em et al.}'s original work investigated the spatial organization and autophagy in angiogenesis, here we focus on characterizing the dynamics and drivers governing these phenotypic transitions. For consistency with Lee {\em et al.}, we retain the original cell-state annotations glycolytic, stalk, and tip; rather than the usual  phalanx, stalk, and tip terminology present in the literature on angiogenesis.

Our analysis begins with $15063$ cells recovered from the fibroblast co-culture condition after performing quality controls and filtering steps (see~\smrefnocite{sec:EC_and_phenotypic}). The cell-cycle-corrected and normalized expression matrix was used to manually annotate cell types, recovering $2372$ ECs, the same number of ECs identified in the original work by Lee {\em et al.} (2024)~\cite{lee2024angiogenesis}. We use the raw transcript counts for each sampled cell to obtain the $1940$ highly variable genes that unambiguously match the authors' original dataset.  
These counts are normalized independently for each cell, log-transformed, and standardized gene by gene, thereby representing each cell as a data point in a common expression space. We proceed by constructing nearest-neighbor graphs in principal component (PC) space and use Leiden clustering to partition the EC population into seven clusters. We impose internal consistency by iteratively removing cells whose Leiden assignment does not coincide with their nearest cluster centroid. This filtering stage retains $1896$ ECs across the $1857$-gene panel, yielding the final set of phenotype-annotated ECs used for downstream analysis. The resulting clusters are grouped into three phenotypic classes: Glycolytic ($\textrm{G}$), Stalk ($\textrm{S}$), and Tip ($\textrm{T}$), containing $1320$, $344$, and $232$ cells, respectively (see~\smrefnocite{sec:EC_annotation}). 

In parallel, we conduct RNA velocity inference~\cite{la2018rna,bergen2020generalizing} on the $1896$ phenotype-annotated ECs  (for a detailed description of our workflow, see~\smrefnocite{subsec:rna-velo-pipeline_SM}). RNA velocity methods estimate the local derivative (rate of change) of the population of spliced transcripts for a given gene from a static ensemble of cells. The method is based on an internal kinetic model of transcript populations that describes the transcription-splicing-degradation dynamics of genes (for details, see~\smrefnocite{subsubsec:rna-velocity-overview}). After aligning the expression and velocity data, we retain only genes exhibiting non-zero velocity information. This procedure results in a final dataset composed of our $M=1896$ cells now across a subset of $N=1873$ supported genes (see~\smrefnocite{sec:rna-velocity-fit}). At the end of this process, we retrieve two data $N\times M$ real matrices $Z=(z_{ai})$ and $W=(w_{ai})$, whose elements respectively represent the centered and standardized spliced-transcript population density of gene $i$ at cell $a$ and its standardized velocity in units of inverse internal time. We assume these two matrices are related as
\begin{align}
\frac{dZ}{dt}=W\,.\label{eq:matrix_dynamical_model}
\end{align}
Here, the temporal variable $t$ is given in internal time units, which are obtained from the RNA-velocity fit. We introduce a rough calibration method that relates this internal time to physical time (in units of hours) by using data on the typical RNA half-life in human cells~\cite{BNID104747} (see~\smrefnocite{subsubsec:rna-velocity-physical-time} for details). This allows us to transform all kinetic rate estimations from the RNA velocity pipeline to units of $1/(\textrm{hours})$ via the obtained scaling factor $\lambda\simeq 0.38$ hours$^{-1}$.

Finally, we combine the outputs of our two data pipelines above by appending the obtained phenotypic labels to the $M=1896$ ECs represented in the aligned expression and velocity matrices. Then, we use the resulting common panel of $N=1873$ genes with valid RNA-velocity estimates to recompute the phenotype-centroids $(z_{\textrm{G}},z_{\textrm{S}},z_{\textrm{T}})$. In the subsequent analysis, we represent our processed data in a high-dimensional cloud of data points (cells) in expression space, each equipped with a velocity vector. The three phenotypic centroids will help us define the empirical expression patterns to be used as the stored memories in the Hopfield-network description.

\section{NONRECIPROCAL HOPFIELD NET APPROACH}
\label{sec:nonrep_hopfield}

In this section, we build an effective model of the phenotype dynamics observed in the angiogenesis experiments of Lee {\em et al.} (2024)~\cite{lee2024angiogenesis} using a nonreciprocal Hopfield network. In their standard form, Hopfield nets~\cite{hopfield1982neural, hertz1991introduction} use symmetric couplings, and for the conventional asynchronous dynamics, this reciprocity allows for the construction of an energy function, where the dynamics are viewed as relaxation on a high-dimensional potential landscape toward a set of stored memory states. This framework naturally accommodates the interpretation of cellular phenotypes as attractors but cannot sustain directed steady-state currents.

Here, we relax the reciprocity constraint and allow asymmetric couplings, meaning that the effective influence exerted by degree of freedom $j$ on $i$ need not equal that exerted in the reverse direction. The standard Hopfield energy construction is then generally unavailable\footnote{Nonreciprocal dynamics can nonetheless be embedded in a higher-dimensional constrained Hamiltonian system by introducing auxiliary degrees of freedom. For recent literature on these methods, see~\cite{shi2026hamiltonian}.}. Consequently, the dynamics do not follow the gradient of a scalar potential function. Instead, the stationary distribution may support nonzero probability currents, while the deterministic mean-field dynamics can admit cycles and other time-dependent behavior. 
In what follows, we provide a strategy to infer the network of nonreciprocal interactions governing our system of interest. Importantly, these inferred couplings provide a coarse-grained model of the phenotypic dynamics of the angiogenic cell population; however, they do not directly establish microscopic nonreciprocal interactions between the underlying molecular regulatory elements.

Asymmetric recurrent networks have been present in the statistical physics literature. For example, they were used to encode temporal associations and cycles of stored patterns~\cite{sompolinsky1986temporal} or as a route to chaos~\cite{sompolinsky1988chaos}. More recently, nonreciprocal interactions have become ubiquitous in active matter physics (e.g., see~\cite{you2020nonreciprocity}), while other work has focused on the role of nonreciprocity as a mechanism for time-dependent phases and associated phase transitions~\cite{fruchart2021non}. Within the Hopfield setting, asymmetric couplings have also been shown to induce switching behavior and limit-cycles among stored memory states~\cite{xue2025critical}.

In this approach, we encode the phenotypic states of endothelial cells relevant to angiogenesis (glycolytic, stalk, and tip)  as associative memories. We analyze how a minimal nonreciprocal model reproduces the organization and directionality of the observed phenotypic flow in gene expression space and provide a platform to bridge the gap between a phenomenological transcriptomic description and more mechanistic theories of differentiation and phenotype switching that are generalizable to other sequencing data. Extensions and future research directions are discussed in Sec.~\ref{sec:results}.

\subsection{Set-up}
\label{subsubsec:set-up}

In Sec.~\ref{sec:data}, we distilled the data of Lee {\em et al.} (2024)~\cite{lee2024angiogenesis} into an expression matrix, $Z=(z_{ai})$, and an aligned RNA-velocity matrix, $W=(w_{ai})$, both containing $M=1896$ cells and the same $N=1873$ dynamically supported genes. We use index notation $a\in\{1,\ldots,M\}$ and $i\in\{1,\ldots,N\}$ to label cells and genes, respectively. Therefore, each cell is regarded as an observed realization of an $N$-dimensional system.

As a first coarse-graining, we discretize the standardized expression of every gene according to\footnote{We use $\operatorname{sign}(x):=+1$ for $x\geq 0$ and $\operatorname{sign}(x)=-1$ for $x<0$.}
\begin{align}
    S=\operatorname{sign}(Z)\in\{-1,+1\}^{M\times N}\,.\label{eq:Z-to-spins}
\end{align}
The $-1$ and $+1$ values respectively represent expressions below and above the mean of each gene, averaged across the reference EC population. Thus, each row in $S=(S_{ai})$ is a spin-like vector, $\bS_a\in\{-1,+1\}^N$, which represents the {\em mesostate} of cell $a$.  We denote with $z_{\mu}$ the standardized expression-space centroids of phenotype $\mu\in\Gamma:=\{\textrm{G},\textrm{S},\textrm{T}\}$, corresponding to the glycolytic, stalk, and tip states. We then associate  each centroid with their corresponding empirical memory pattern
\begin{align}
    \bximu :=\operatorname{sign}(z_{\mu})\,.\label{eq:centroid-to-memory-states}
\end{align}
These patterns are not orthogonal in general. 
Correlations between stored patterns are known to influence the collective retrieval properties of associative-memory networks~\cite{agliari2013parallel}. Here, the correlations arise naturally from the experimentally determined phenotypic states. The similarity between a generic cellular configuration $\bS$ and phenotype $\mu\in\Gamma$ is quantified by the overlap
\begin{align}
    m_{\mu}(\bS):=\frac{1}{N}\bximuT \bS\,.\label{eq:overlap-def}
\end{align}
The three overlaps encoded in $\bm^{\top}:=(\mP,\mS,\mT)$ provide a reduced phenotypic coordinate system. Below, we use this reduced representation to compare the empirical and the Hopfield-modeled dynamics. The empirical correlations between memory patterns and the appropriate gauge transformations applied to our set-up are detailed~\smrefnocite{sec:nonrep_hopfield_sm}.

\vspace{-1em}
\paragraph*{{\bf Interactions}.} 

In the absence of external forcing, the local field acting on the system state is given by
\begin{align}
    \bh(\bS)=J\bS,\label{eq:h} 
\end{align}
where $J\in\Reals^{N\times N}$ encodes the effective interactions between genes in their spin-like expression-space.
In our approach, $J$ is typically asymmetric (nonreciprocal) and is the main object of our inference approach.

\begin{figure}
    \centering
    \includegraphics[width=\linewidth]{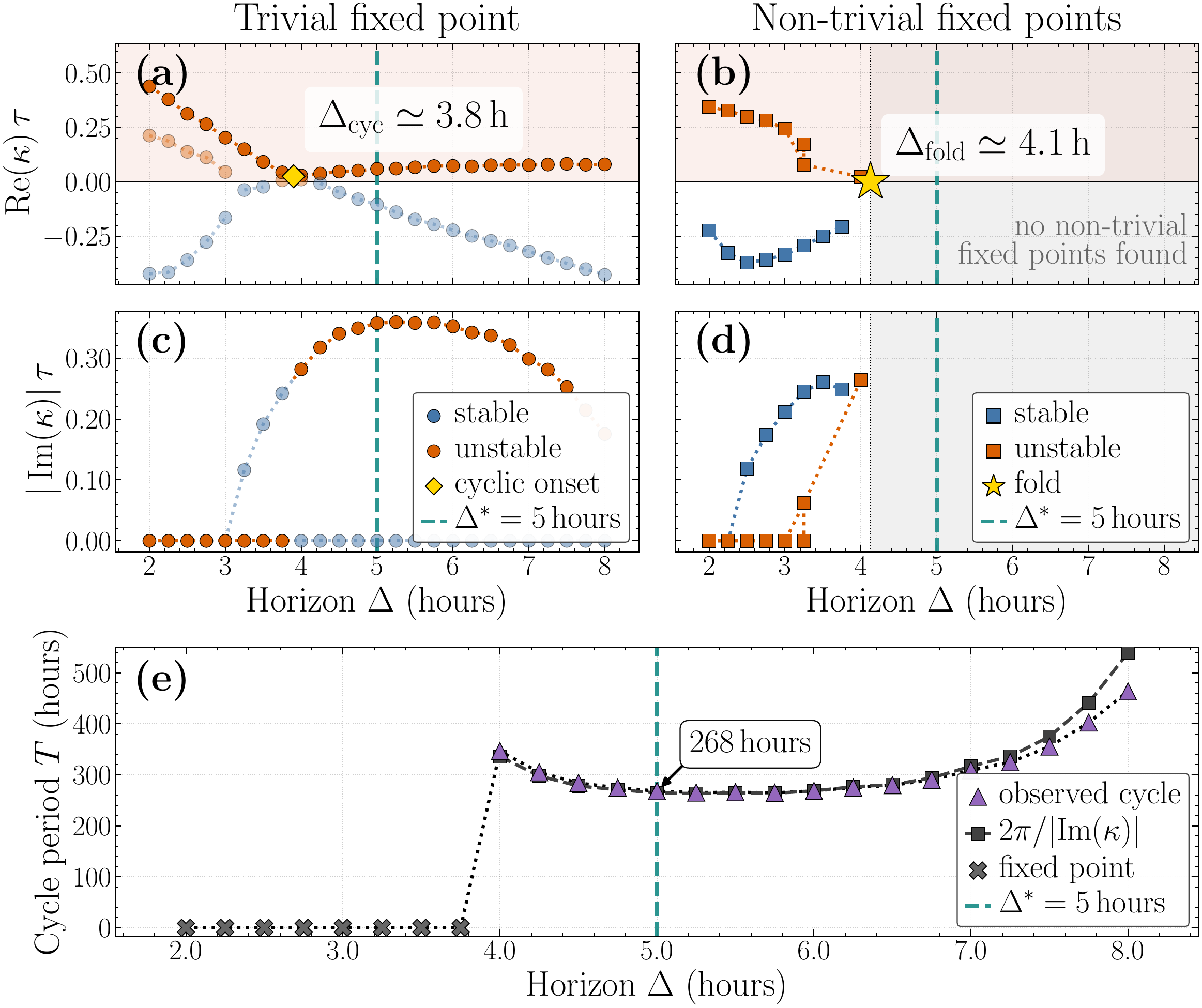}
    \caption{{\bf Spectral analysis of the DMFT across horizon values.} 
    Plots \textbf{(a)}-\textbf{(d)} show the dimensionless eigenvalues real (top panel) and imaginary (middle panel) parts $\kappa\tau$ evaluated at the trivial ($\bm^*_0=\boldsymbol{0}$, left column) and non-trivial ($\bm^*\neq \boldsymbol{0}$, right column) fixed points; for our range of horizon values $\hor$. Plots \textbf{(a, c)} show the three eigenvalues and their dependence on $\hor$ (with subdominant eigenvalues in faded markers), while \textbf{(b, d)} show only the dominant (maximum real-valued) eigenvalue behavior for all the non-trivial fixed points. Panel \textbf{(e)} summarizes the corresponding long-time behavior observed on simulated DMFT trajectories: gray crosses denote convergence to a fixed point, and purple triangles show the periods of the detected cycles. The vertical green dashed line marks the horizon value selected for further analysis, which displays 268 hours cycle.}
    \label{fig:FIG2}
\end{figure}

We introduce an asymmetric {\em mixing matrix}, $\mix\in\Reals^{|\Gamma|\times |\Gamma|}$, which allows us to relate the interaction matrix $J$ to pattern-to-pattern couplings,
\begin{align}
    J=\frac{1}{N}\sum_{\mu,\nu}\bximu\mix_{\mu\nu}{\bxinu}^{\intercal}\,,
    \label{eq:JHop}
\end{align}
and setting $J_{ii}=0$ afterward.
In our system, this introduces a set of $3\times 3$ unrestricted parameters, $\{\mix_{\mu\nu}\}$, which will be inferred below using our processed data. The nonreciprocal nature of the system is now linked to the structure of the matrix $\mix$.  In particular, when $\operatorname{asym}(\mix)=0$, 
then 
$J_{ij}=J_{ji}$ for all $i,j$. Below, we discuss the implications of the nonsymmetric component of our inferred matrix $\mix$. 

Overall, Eq.~\eqref{eq:JHop} generalizes the standard Hebbian construction by allowing a full mixing matrix in the subspace spanned by the empirical phenotype patterns. The symmetric part of $\mix$ modifies the reciprocal interactions within this subspace, whereas its antisymmetric part generically produces asymmetric couplings, thereby breaking detailed balance and permitting directed probability currents between phenotype-associated regions of state space.

\begin{figure}
    \centering
    \includegraphics[width=\linewidth]{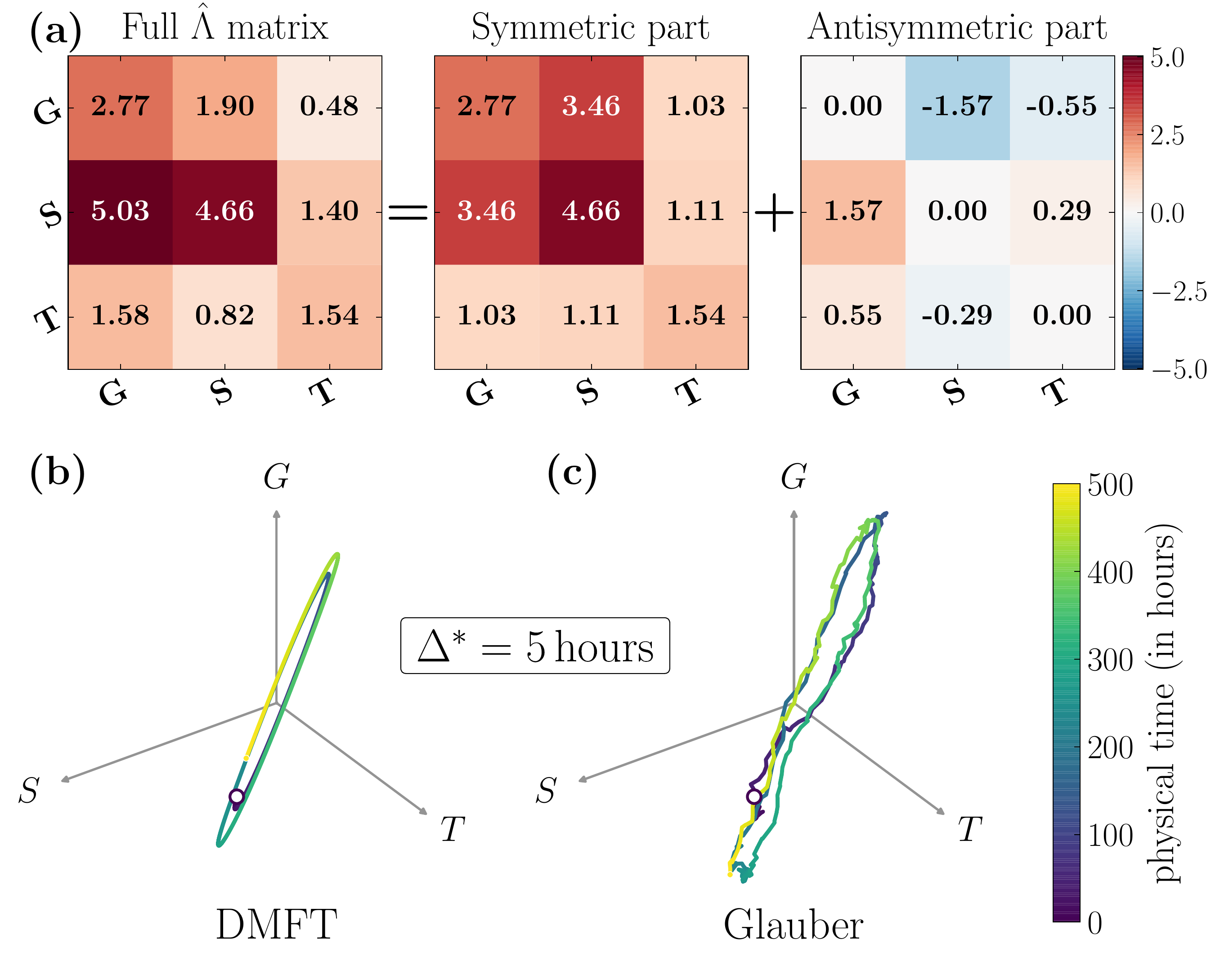}
    \caption{{\bf Fit \& dynamics.} Panel {\bf (a)} shows the result for fit of the mixing matrix $\hat{\mix}$ and its decomposition in symmetric and antisymmetric parts for the choice of extrapolation horizon $\hor^*=5$ hours. Panels {\bf (b-c)} show the DMFT and Glauber trajectories projected on the overlap space. Trajectories start at the marked empty node in blue and evolve over physical time up to 500 hours. Note the overlaps are not truly orthogonal (see~\smrefnocite{subsec:model-outputs}). Overlap axes are here depicted orthogonally only for simplification purposes.}
    \label{fig:FIG3}
\end{figure}

\subsection{Dynamical mean-field reduction}
\label{subsec:dmft}

We evolve the nonreciprocal Hopfield system using continuous-time asynchronous Glauber dynamics (see~\smrefnocite{subsec:sm_glauber} for details), where spins are updated one at a time at a rate determined by their local field, Eq.~\eqref{eq:h}, and the time-parameter $\tau$, which sets the update timescale. This timescale is also set by the kinetic parameter estimations outputted by the RNA velocity algorithm (see~\smrefnocite{subsubec:calibration_glauber_tau} for details) giving back an estimate (in physical units) of $\tau\simeq 15.0$ hours. We also note that the usual inverse-temperature factor is absorbed into the local field. This dynamics generates an ensemble of stochastic trajectories in the space $\bS(t)\in\{-1,+1\}^N$. We resolve a statistic of these trajectories at the phenotypic level by studying the ensemble-averaged overlaps. We close these averaged dynamics at the level of $\bm(t)$ by following a dynamical mean-field theory (DMFT)  approximation~\cite{crisanti1987dynamics,crisanti1988dynamics} (see~\smrefnocite{subsec:dmft_sm} for details). The resulting dynamical system reads as\footnote{We use Hadamard product notation, $\circ$, between vectors of the same space, such that the resulting components follow: $(x\circ y)_i =  x_iy_i$.}
\begin{align}
    \tau\frac{d}{dt}\bm &= -\bm+ \mathfrak{H}\left(\brho\circ \tanh(\mathfrak{H}
    ^{\intercal}\mix \bm )\right)\,,\label{eq:DMFT-m-overlaps-vs-group-fields-main-text}   
\end{align}
with the definition $\brho := (1/4)H_4 \bQ$,
where $H_4$ is the Hadamard-Walsh matrix of order four, the vector $\bQ$ 
is obtained from the Gram matrix of the three memory states, and $\mathfrak{H}$ is the top $3 \times 4$ block of $H_4$ 
(for more details and a complete derivation, see~\smrefnocite{subsubsec:empirical-corrs}). Importantly, in Eq.~\eqref{eq:DMFT-m-overlaps-vs-group-fields-main-text}, we are using physical time for $t$, which is given in units of hours. We also note that since the empirical memory patterns are correlated, then the three overlaps must generally be interpreted jointly.

\subsection{Fitting $\mix$}
\label{subsec:mix_matrix_fit}

To estimate the entries in the mixing matrix $\mix$, we consider the finite-time evolution of the overlaps obtained from the DMF equations~\eqref{eq:DMFT-m-overlaps-vs-group-fields-main-text} and compare them against the changes obtained by integrating the expression $Z$ using the RNA velocity estimates $W$. This fit uses the whole valid panel of $N=1873$ genes and the empirical memory patterns introduced in~\eqref{eq:centroid-to-memory-states}. We introduce a new temporal parameter termed {\em horizon}, which we denote by $\hor$, given in units of hours. Below, we use this parameter to shift the expression matrix $Z(0)\to Z(\hor)$. As we discuss next, the horizon parameter is varied to assess the sensitivity of the inference and yields distinct estimates for the mixing matrix at each horizon value, $\hat{\mix}(\hor)$.

\begin{figure}
    \centering
    \includegraphics[width=\linewidth]{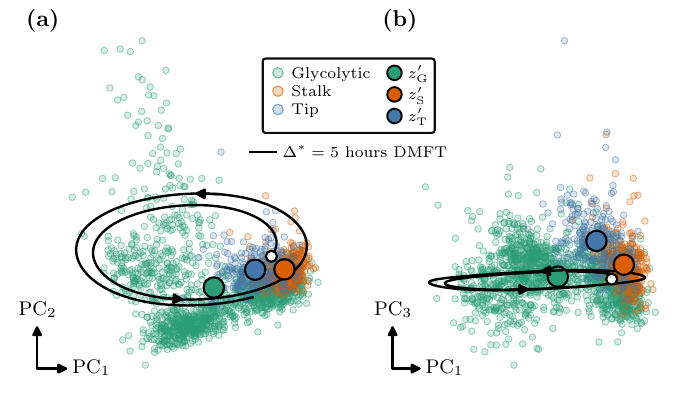}
    \caption{\textbf{Mean-field dynamics in principal-component space.}
    Panels {\bf (a)} and {\bf (b)} show the endothelial-cell expression profiles projected onto the PC1--PC2 and PC1--PC3 planes, respectively. Small circles denote individual cells, colored by phenotype, while the larger outlined circles indicate the corresponding PC-projected phenotypic centroids $z'_{\mathrm{G}}$, $z'_{\mathrm{S}}$, and $z'_{\mathrm{T}}$. The black curve shows the DMFT trajectory obtained for the selected horizon value, $\hor^*=5\,\textrm{hours}$. The open circle marks the initial condition and the arrowheads indicate the direction of motion.}
    \label{fig:FIG4}
\end{figure}

For statistical purposes, the set of $M=1896$ ECs is randomly partitioned into a {\em training} and a {\em test} subsets of $1516$ and $380$ ECs, respectively. The gene means, standard deviations, and phenotype centroids are recomputed for the training subset. These quantities, and the resulting memory patterns, act as references for the fit and remain fixed during the construction of the optimization targets determined by each $\hor$ value.

In the training set, consider the standardized expression and RNA velocity matrices $Z$ and $W$, respectively. In the original RNA velocity fit, $W$ is expressed in units of inverse internal time; thus, its physical-time counterpart is $\lambda W$. We then extrapolate expressions as $Z+\lambda \hor W$ and apply the threshold in~\eqref{eq:Z-to-spins}, which gives $\bS(\hor)$. We define the target overlaps as 
\begin{align}
m_{\mu}^{\hor}:=\frac{1}{N}\bximuT \bS(\hor)\,.
\label{eq:target-overlaps}
\end{align}
No additional standardization is applied to the extrapolated expressions. Next, consider a candidate mixing matrix, $\mix$, and use it to numerically integrate the DMFT equations~\eqref{eq:DMFT-m-overlaps-vs-group-fields-main-text} with initial conditions $\bm^0$ and over the physical time duration $\hor$. We denote this deterministic flow by $\Phi_{\mix}^{\hor}(\bm^0)$, which is the model-predicted endpoint. 
We then estimate the best common mixing matrix $\hat{\mix}(\hor)$ for all cells by nonlinear least-squares between the endpoints $\{\Phi_{\mix}^{\hor}(\bm_a)\}$ and the respective extrapolated targets $\{\bm_a^{\hor}\}$, for all $a$ in the training set. Thus, the unrestricted fit estimates nine mixing entries, while $\tau$ remains fixed. The loss definition, numerical implementation, and statistical diagnostics are detailed in~\smrefnocite{subsec:SM-mixing-matrix-fit}.

We estimate mixing matrices across a range of $\hor$ values between $2-8$ hours. This choice is motivated by empirical observations on how angiogenic ECs reorganize their functional state. For example, time-lapse imaging shows that ECs physically overtake one another, thus replacing tip cells approximately every $3-5$ hours~\cite{jakobsson2010endothelial}. In mice, it has been observed that the Vegf-Dll4/Notch circuit exhibits fluctuations directly measured on the Dll4 expression with a timescale of $5-6$ hours~\cite{ubezio2016synchronization}, which drive EC phenotypic reorganization. In Fig.~\ref{fig:FIG3}, we report the results obtained for $\hor^*=5\,\textrm{hours}$, which display an interesting cyclic behavior discussed in Sec.~\ref{sec:results}. 

Finally, we assessed the predictive value of our nonreciprocal Hopfield model by testing its accuracy against a constraint reciprocal version and a simple persistence counterpart (see~\smrefnocite{subsec:SM-mixing-matrix-fit} for details). At each horizon value, we evaluate both our nonreciprocal model and its restricted/reciprocal counterpart against the same 380 held-out (test) cell targets. Our results show that the nonreciprocal fit, $\hat{\mix}$, is the best performing model. To this effect, in~\smrefnocite{subsec:SM-mixing-matrix-fit}, we consider the statistical metrics RMSE and the determination coefficient ($R^2$), and perform an $F-$test to compare both fits, accounting for the different number of parameters.  Performance indicators strongly increase in favor of our nonreciprocal fit, showing higher statistical significance at the higher end of our $\hor$ range.

\subsection{Results}
\label{sec:results}

After the fitting procedure, we numerically determine the fixed points of Eq.~\eqref{eq:DMFT-m-overlaps-vs-group-fields-main-text} using our obtained fits for the mixing matrices $\hat{\Lambda} (\hor)$ in the range of horizon values from $2-8$ hours.  In Fig.~\ref{fig:FIG2}a-d, we study the set of eigenvalues, $\{\kappa\}$, of the fixed points found at each $\hor$ value. The top and middle panels show the respective real and imaginary values. The left column (a,c) concerns the trivial point (i.e., the origin $\bm^*_0=\0$), and the right column (b,d) involves the non-trivial fixed points ($\bm^*\neq \0$). The spectrum at the origin exhibits three distinct events as we increase the horizon parameter $\hor$. First, at around $\hor=3.1\,\textrm{hours}$, the subdominant eigenvalues (shown in faded markers) collide, and a transition from two reals into a complex conjugate pair occurs. The resulting shared real part remains negative until crossing over to the positive (unstable) region at a $\hor$ value close to but smaller than $\hor_{\textrm{cyc}}$, which is shown in (a) as a yellow diamond. At $\hor_{\textrm{cyc}}\simeq 3.9\,\textrm{hours}$, the now unstable complex pair overtakes the remaining purely real eigenvalue. Thus, for $\hor\gtrsim \hor_{\textrm{cyc}}$, the trivial fixed point is dominated by a complex unstable mode with a characteristic oscillatory frequency governed by its imaginary part. 

For the spectrum of the non-trivial fixed points, we plot only the data (in squares) of the eigenvalues with the maximum (red) and minimum (blue) real parts. Here we identify three relevant regimes. First, the region of horizon values between $2-3.1$ hours is where both stable and unstable non-trivial fixed points coexist. Here, the stable ones correspond to memory retrieval attractors, and the unstable points are saddles. Second, the region of $\hor$ between $3.1-4.0$ hours, in which no saddle (unstable) modes are found. Within this narrow region, retrieved memories cannot reliably be distinguished from each other, since saddle-like points no longer separate them. This facilitates directed transitions between memory states (see~\cite{antonioinprep} for a detailed explanation of this phenomenon). And third, above the horizon value of $\hor_{\textrm{fold}}$, where we no longer find non-trivial fixed points. This suggests that the non-trivial fixed points undergo a fold-type transition, which is consistent with simpler theoretical toy-models~\cite{xue2025critical,antonioinprep}. 

In Fig.~\ref{fig:FIG2}e, we monitor simulations of the DMFT equations for the $\hor$ values and characterize their late-time behavior to check for the presence of persistent cycles (for more details, see~\smrefnocite{subsec:model-outputs}). The plotted gray crosses (displayed on the horizontal axis for convention only) correspond to $\hor$ values for which no persistent oscillations are found, while purple triangles mark the observed periods for the $\hor$ values where cycles are observed. The observed periods approach the values predicted from the unstable cyclic mode of the trivial fixed point (black boxes), using $2\pi/\lvert \textrm{Im}(\kappa)\rvert$, where $\textrm{Im}(\kappa)$ here is the imaginary part of said mode found for $\hor>\hor_{\textrm{cyc}}$.

In Fig.~\ref{fig:FIG3}a, we show the symmetric and antisymmetric parts of the nonreciprocal fitted matrix $\hat{\mix}$ obtained at the selected horizon of $\hor^*=5\,\mathrm{hours}$. The presence of a nonzero antisymmetric component enables cycles to govern the deterministic dynamics. Its DMFT and stochastic Glauber trajectories are shown in Fig.~\ref{fig:FIG3}b,c. The stochastic trajectories closely follow the DMFT predictions when cycles are observed. However, when the deterministic trajectories tend toward a fixed point, the Glauber fluctuations show exchanges between fixed states driven by noise (for more details, see~\smrefnocite{subsec:model-outputs}). 

Fig.~\ref{fig:FIG4} displays the DMFT trajectory found with $\hor^*=5\,\textrm{hours}$ projected onto the principal-component coordinates of the expression data. It includes all the EC data points and the PC-projected phenotype-centroids, $\{z'_{\mu}\}$ all color coded with their phenotype labels, with the predicted (deterministic) motion superposed. In this depiction, the randomly chosen initial cell (which provides the initial conditions for the DMFT trajectory) belongs to the Tip cluster. The DMFT trajectory then flows into the Glycolytic region to return to the Stalk-Tip region, which appears closely packed together in both shown projections, thus closing the cycle. This flow represents an effective progression of a typical cell through expression-similarity space.

\section{Discussion}
\label{sec:discussion}

In this work, we fitted a nonreciprocal Hopfield network to scRNA-seq data on endothelial cells under certain experimental culture conditions. We have drawn phenotypic annotations from the authors of the experimental paper~\cite{lee2024angiogenesis}, and used the raw sequencing data to produce RNA velocity estimates. We used the processed data to fit a family of mixing matrices by varying a temporal horizon parameter, $\hor$, whose values are chosen based on empirical observations. We have demonstrated that our method improves the predictive capacity of a symmetric associative memory approach and allows us to predict possible long-time cycles present in the cellular program.

In our set-up, we did not consider the influence of external forcing. Here, we assumed that the influence of signals from the co-cultured fibroblasts remains present in the data and is therefore absorbed into the inferred couplings. However, in the real-world, fluctuations in the signal can arise, as well as other conflicting bimolecular signals. A potential strategy to incorporate these exogenous forces comes from introducing explicit external fields added to the local field $\bh$. These fields can be tailored to specific genes (e.g., for specific receptor proteins). 
In particular, this approach would allow for the exploration of  sequential perturbations. For example, after fitting a given dataset, one could numerically explore how interventions in the order of a developmental program may change the outcome. This connects naturally to Yamanaka's experiments, which showed that the sequential delivery of factors conduces to effective reprogramming~\cite{takahashi2006induction,liu2013sequential}. Nonreciprocal interactions provide an ideal framework for studying directed (and therefore nonequilibrium) regulatory effects. This framework could also be used to explore epistatic effects, for example, by comparing paired gene perturbations versus individual ones and running the inferred dynamics over time. 

Finally, another promising direction to extend our findings arises when confronting the need for more mechanistic and causality-based modeling in the inference of gene regulatory networks~\cite{maizels2026gene}. A useful extension would consist of constraining the gene-to-gene interaction matrix $J$ so that only real molecular interactions are allowed (e.g., via transcription factor or equivalent protein-promoter systems). This would effectively mask our matrix defined in Eq.~\eqref{eq:JHop} to give $\tilde{J}=A\circ J$, where $A$ encodes the possible (directed) factor-target annotations available. This modified network would require refitting and reconsideration of the mean-field dynamical reduction scheme. It is also possible to extend the Hopfield formulation to a continuous-state, such as $\bz=(z_1,\ldots,z_N)$, where the information for each gene is stored in terms of standardized expression (rather than spin-like variables). Here, we could study dynamical effects on expression amplitudes and explore potential regulatory pathways leading to resonant states, which could be linked to cell dysfunction.

\begin{acknowledgments}

The authors thank Jong Kyoung Kim, Bum Suk Kim for kindly providing data used in this study. JP acknowledges support from Taighde \'Eireann -- Research Ireland under the Government of Ireland Postdoctoral Fellowship Programme GOIPD/2025/1111. JP thanks Prof. Kevin Mitchell for fruitful discussions. The authors acknowledge the computer resources (Picasso Supercomputer), technical expertise, and assistance provided by the SCBI (Supercomputing and Bioinformatics) center of the University of M\'alaga. CP acknowledges support from the NIH under R35GM149261. This publication does not represent official views of the NIH.
\end{acknowledgments}

\bibliographystyle{apsrev4-2}
\bibliography{bibliography}

\clearpage

\newcommand\newstuff[1]{#1}

\ifarxiv 
\let\addcontentsline\savedaddcontentsline

\pdfstringdefDisableCommands{%
  \renewcommand{\bm}[1]{#1}%
}

\hypersetup{
  colorlinks,
  linktocpage=true,
  linkcolor={red!50!black},
  citecolor={blue!50!black},
  urlcolor={blue!80!black}
}

\setcounter{tocdepth}{2}

\newcommand\ak[1]{{\color{red} AK: #1}}

\newcommand{\myeqref}[1]{(\ref{#1})}

\newenvironment{newstuffenv}
    {\color{blue}}
    {}

\global\long\def\bg{\boldsymbol{g}}

\newcounter{myboxcounter}[section]
\renewcommand{\themyboxcounter}{\thesection.\arabic{myboxcounter}}
\def\thesection{\arabic{section}}

\newtcolorbox[auto counter, number within=section]{mybox}[2][]{colback=gray!20, colframe=white, fonttitle=\bfseries, title=Box~\thetcbcounter: #2,#1}

\newcommand\figurefontcmd{\fontsize{10pt}{12bp}\selectfont}

\makeatletter
\renewcommand\section{\@startsection{section}{1}{\z@}%
                                  {-3.5ex \@plus -1ex \@minus -.2ex}%
                                  {2.3ex \@plus.2ex}%
                                  {\fontsize{12pt}{12bp}\selectfont\bfseries\centering}}
\renewcommand\subsection{\@startsection{subsection}{2}{\z@}%
                                  {-3.5ex \@plus -1ex \@minus -.2ex}%
                                  {2.3ex \@plus.2ex}%
                                  {\fontsize{11.5pt}{12bp}\selectfont\bfseries\centering}}

\renewcommand\subsubsection{\@startsection{subsubsection}{3}{\z@}%
                                  {-3.5ex \@plus -1ex \@minus -.2ex}%
                                  {2.3ex \@plus.2ex}%
                                  {\fontsize{11.5pt}{12bp}\selectfont\em\centering}}
\makeatother

\sloppy
\allowdisplaybreaks
\raggedbottom

\onecolumngrid
\setcounter{equation}{0}

\renewcommand{\theequation}{S\arabic{section}.\arabic{equation}}
\renewcommand{\thefigure}{S\arabic{figure}}
\renewcommand{\thesection}{S\arabic{section}}

\linespread{0.8}

\onecolumngrid
\setcounter{section}{0}

\begin{center}

{\fontsize{15pt}{22bp}\bfseries Supplemental Material: Nonreciprocal Hopfield nets for angiogenesis}

\vspace{15pt}

{\fontsize{10pt}{15pt}\selectfont Jordi Pi\~nero, Krishna Chaitanya Ginne, Antonio Alves Ducci, Ra\'ul Monta\~nez, and Carlo Piermarocchi\par}

\end{center}

\fontsize{11pt}{14.5bp}\selectfont
\tableofcontents{}

\fontsize{11pt}{22bp}\selectfont

\newpage

\section{Data Pipeline}
\label{sec:pipeline}

The data that we use throughout this work is single-cell RNA sequencing data (scRNA-seq). We base all our analysis on the data obtained in the experiment performed by Lee {\em et al.} (2024)~\cite{lee2024angiogenesis}. In these experiments, the authors isolated single cells using a droplet-based microfluidic platform, where individual cells were encapsulated with barcoded gel beads in nanoliter droplets on a microfluidic chip. These droplets act as independent compartments in which transcripts are captured and labeled, enabling downstream sequencing to resolve cell-specific expression profiles. The NCBI reference for these experiments is found in \href{https://www.ncbi.nlm.nih.gov/bioproject/PRJNA931762}{PRJNA931762}.

Lee {\em et al.}'s (2024)~\cite{lee2024angiogenesis} analysis involves two scRNA-seq datasets collected from two samples, both containing human umbilical cord endothelial cells (HUVECs). The first sample (hereby termed W-sample) combines HUVECs {\em with} lung fibroblasts (LFs), while the second sample (hereby termed WO-sample) was processed {\em without} any LFs. Fibroblasts provide the pro-angiogenic signals that induce sprouting. Accordingly, our downstream analysis focuses only on the W-sample. The relevant NCBI database for the W-sample is available under reference \href{https://trace.ncbi.nlm.nih.gov/Traces/?run=SRR23344353}{SRR23344353}. The libraries we use were prepared with the \texttt{Chromium 10X Genomics} platform and processed by Lee {\em et al.} (2024)~\cite{lee2024angiogenesis} with \texttt{Cell Ranger} to classify the barcodes~\cite{lee2024angiogenesis}. The processed EC assignments used below were kindly provided by Bum Suk Kim and Jong Kyoung Kim. 

In what follows, we divide our analysis into three parts. In SM-\ref{sec:EC_and_phenotypic}, we outline the initial EC filtering, whose output forms the basis of our analysis. In SM-\ref{sec:EC_annotation}, we recompute and match the authors' original ECs classification into the three relevant phenotypes in the context of angiogenesis: {\em Glycolytic} (G), {\em Stalk} (S), and {\em Tip} (T). Finally, in SM-\ref{subsec:rna-velo-pipeline_SM}, we use the original scRNA-seq data to compute the RNA velocity across the annotated ECs.

\subsection{Lee {\em et al.}'s (2024) original EC Filtering \& Phenotypic Classification}
\label{sec:EC_and_phenotypic}

Authors in Lee et al. (2024)~\cite{lee2024angiogenesis} preprocessed the raw scRNA-seq data before any of the analysis reported in this work. Our study begins with their preprocessed endothelial cell (EC) dataset. For completeness, we briefly summarize the main preprocessing steps below.

\paragraph*{Quality controls ---} The original dataset comprised $2{,}084{,}239$ and $2{,}351{,}192$ barcoded droplets from the W and WO conditions, respectively, for a total of $4{,}435{,}431$ droplets profiled across $36{,}601$ genes. As is typical for droplet-based scRNA-seq, only a small fraction of these droplets corresponded to genuine cells.

The first quality-control stage removed droplets with fewer than approximately $160$ total counts, filtered cells with mitochondrial content exceeding $10\%$, and discarded low-quality cells with fewer than $1000$ total transcript counts. After these filtering steps, $15{,}063$ W cells and $13{,}621$ WO cells remained.

A second quality-control stage normalized gene expression using the \texttt{scran} package, identified highly variable genes (HVGs), and corrected cell-cycle effects using predefined S- and G2/M-phase marker genes through the \texttt{Seurat} workflow. Under the authors' default parameters, this resulted in $2777$ HVGs for downstream analysis.

\paragraph*{EC Classification --- }Using the corrected expression matrix, the authors projected the data onto the first $30$ principal components, constructed a nearest-neighbor graph, and performed graph-based clustering ($k=10$, resolution $0.5$). The resulting clusters were manually annotated to identify endothelial cells, yielding $2372$ ECs in the W-sample and $12{,}186$ ECs in the WO-sample.

The selected ECs were subsequently reprocessed by repeating normalization, HVG selection, cell-cycle correction, dimensionality reduction, and graph clustering on the EC subset alone. Because the WO-sample contained substantially more ECs than the W condition, the authors clustered the WO ECs and sampled at most $250$ cells from each cluster, producing a representative subset of $2321$ WO ECs. Finally, the W and WO datasets were normalized separately, a common set of $2000$ integration features was selected, and \texttt{Seurat}'s \texttt{FindIntegrationAnchors} and \texttt{IntegrateData} procedures were used to perform batch correction and integration (see~\cite{lee2024angiogenesis} Methods Section, under ``Cell sampling for batch correction'').

\subsection{Clustering \& Phenotype Annotation}
\label{sec:EC_annotation}

\paragraph*{Clustering --- }Our analysis begins with the EC subset identified and processed by Lee {\em et al.} (2024)~\cite{lee2024angiogenesis} and described above. The author's final integrated dataset contains $4693$ ECs, comprising $2372$ cells from the W-sample and $2321$ cells from the WO-sample, and is represented over a common panel of $2000$ integration features. The rest of the analysis presented below focuses on the W-sample. 

We return to the corresponding raw-count expression matrix while using the authors' processed dataset to determine the set of cells and genes retained after their preprocessing. Gene identifiers in the raw-count matrix are first harmonized with the gene symbols represented in the authors' processed reference. Of the $2000$ integration features in the processed dataset, $1940$ can be unambiguously matched to genes in the raw-count matrix. We therefore retain these $1940$ genes and restrict the cell population to the $2372$ W-conditioned ECs identified by Lee {\em et al.} Thus, the resulting expression matrix used as input to our analysis contains $2372$ cells and $1940$ genes.

Starting from the raw-count expression matrix described above, we use the \texttt{Scanpy} package to normalize and cluster the ECs in the W-sample. First, we perform library-size normalization using \texttt{sc.pp.normalize\_total} with \texttt{target\_sum} $=10^4$. This rescales the raw gene counts independently for each cell, such that its total expression across all genes equals $10^4$ pseudo-counts. 

For clustering, we first apply \texttt{sc.pp.log1p} to $X$, which transforms each normalized expression value $x\mapsto\ln(1+x)$. We take the resulting log1p-transformed expression matrix and standardize it using \texttt{sc.pp.scale}, which centers the expression of each gene across cells to a mean of zero and scales it to unit variance. We term this log1p-transformed and standardized matrix $Z_0$ for later use. Principal component analysis (PCA) is then performed on $Z_0$ using \texttt{sc.tl.pca}, retaining the first $20$ principal components. From these $20$ PCs, we construct a nearest-neighbor graph using \texttt{sc.pp.neighbors} with \texttt{n\_neighbors}$=10$ and \texttt{n\_pcs}$=20$.

We subsequently perform Leiden clustering on this nearest-neighbor graph using \texttt{sc.tl.leiden} with a resolution parameter of $0.5$. This procedure partitions the $2372$ W-condition ECs into a total of seven Leiden clusters. 

\paragraph*{First Centroid-Consistency Filter --- }Following Leiden clustering, we test the consistency of each cell with its assigned cluster using the same $1940$-gene expression space employed above. For this purpose, we return to the log1p-transformed and standardized matrix $Z_0$, which we recall has dimensions $2372\times1940$.

We start calculating the centroid for each of the seven Leiden clusters by averaging $Z_0$ across all currently retained cells assigned to each cluster, thus producing seven $1940$-dimensional cluster-centroid vectors. For every cell, we calculate its Euclidean distance to each of the seven cluster centroids using \texttt{pairwise\_distances}. The centroid with the smallest Euclidean distance is then identified and compared with the cell's original Leiden cluster assignment. A cell is retained only when its nearest centroid corresponds to its original Leiden cluster; otherwise, the cell is considered inconsistent with its cluster assignment and is thus removed. We note that the original Leiden labels are preserved throughout this procedure and that inconsistent cells are removed rather than reassigned to a different cluster.

Removal of inconsistent cells changes the data composition and, therefore, the centroid of each Leiden cluster. Hence, this consistency test is performed iteratively such that, after each round of filtering, the cluster centroids are recalculated using only the surviving cells, while the standardized expression coordinate system remains fixed. The cell-to-centroid distances are then recomputed, and the consistency criterion is applied again. Iteration continues until no additional inconsistent cells are identified, with a maximum of $10$ iterations.

Starting from $2372$ cells, the first iteration removes $229$ cells, followed by $29$, $9$, $4$, and $1$ additional cells in successive iterations. The procedure converges on the sixth iteration, at which point no further cells are removed. In total, $272$ cells are excluded, and $2100$ W-sample ECs are retained. Hence, the resulting expression matrix after filtering contains $2100$ cells over the same $1940$-gene panel. These retained cells and genes are subsequently used to define the input population for the RNA-velocity pipeline described in~\ref{sec:rna-velocity-fit}. 

As shown below, alignment with the resulting RNA-velocity estimates retains $1896$ cells and $1873$ genes with valid RNA-velocity information. 
In Fig.~\ref{fig:A1}a we plot (in PC space) this final filtered matrix and color all cells according to their final Leiden clustering assignments. Only after this final cell population has been established do we perform the biological phenotype annotation as described next.

\paragraph*{EC Phenotype annotation --- } Following the completion of the RNA-velocity alignment and the final centroid-consistency filter, we assign biological phenotypes to each EC cluster using marker-gene signature scores. For this classification step, we return to the full log-normalized gene-expression matrix rather than restricting the analysis to the $543$ dynamically supported genes retained for the final velocity-based analysis. The expression matrix is aligned to the $1896$ W-sample ECs surviving the final centroid-consistency filter, and their original \texttt{EC\_cluster} labels are transferred to the corresponding cells. This results in a scoring expression matrix containing $1896$ cells and $15{,}945$ unique genes.

Three marker sets are considered, corresponding to Glycolytic, Stalk and Tip EC phenotypes. We use the same Stalk and Tip signatures employed by Lee {\em et al.} (2024)~\cite{lee2024angiogenesis} in their ``Signature scores'' analysis, which are based on established Tip/Stalk markers reported by Chen {\em et al.} (2019)~\cite{chen2019endothelial}. The Glycolytic signature is likewise defined following Lee {\em et al.} (2024)~\cite{lee2024angiogenesis}, using the \texttt{HALLMARK\_GLYCOLYSIS} gene set from the Molecular Signatures Database (MSigDB). Each marker set is first intersected with the $15{,}945$ genes available in our expression matrix. Following this intersection, the maximum common number of markers that can be consistently retained across all three signatures is $18$ genes. We cap each signature set at up to $18$ genes to ensure that the three phenotype scores are computed from equally sized marker sets. For signatures containing more than $18$ available markers, we calculate the expression variance of each marker across the $1896$ cells and retain the $18$ genes with the largest variance. Hence, the final scoring procedure uses $18$ Stalk, $18$ Tip, and $18$ Glycolysis marker genes.

A signature score is then calculated independently for each phenotype and each cell using the \texttt{Scanpy} function \texttt{sc.tl.score\_genes}. For a given marker set, this procedure calculates its average expression in each cell relative to a reference set of genes with comparable expression levels. Consequently, every EC is assigned three scores, denoted \texttt{score\_stalk}, \texttt{score\_tip}, and \texttt{score\_glycolysis}. These cell-level scores are subsequently averaged over all cells belonging to the same Leiden cluster, yielding one Glycolytic, Stalk and Tip score for each of the seven EC clusters.

We next assign a phenotype to each Leiden cluster according to these cluster-level signature scores. If only one of the three signature scores is positive, the cluster is assigned directly to the corresponding phenotype. Otherwise, clusters with a unique maximum score are assigned to the corresponding phenotype. Each cell subsequently inherits the phenotype assigned to its Leiden cluster. This results in $1320$ Glycolytic, $344$ Stalk-like, and $232$ Tip-like ECs, accounting for all $1896$ ECs in the W-sample (see Fig.~\ref{fig:A1}b).

\begin{figure}
    \centering
    \includegraphics[width=0.85\linewidth]{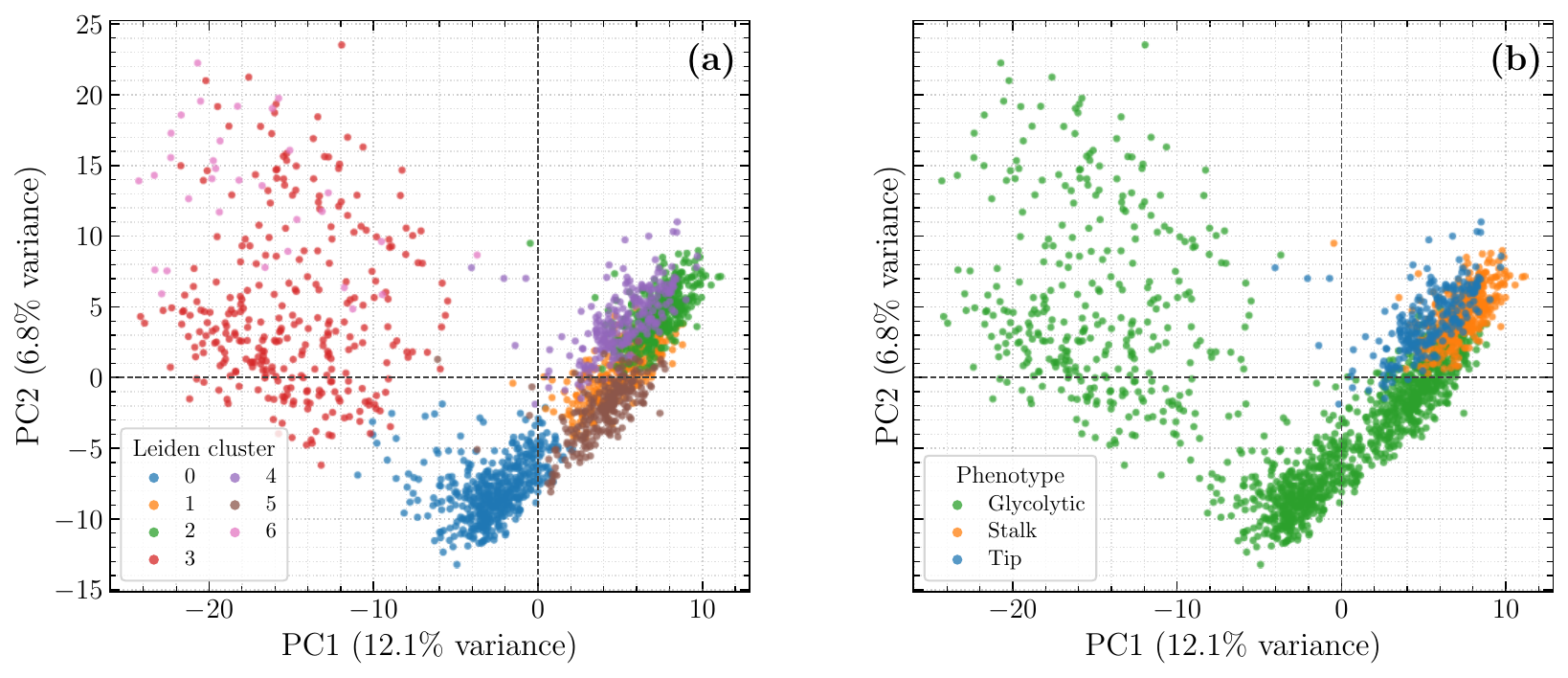}
    \caption{{\textbf{Leiden Clusters \& EC Phenotype Classification.}} Panel {\bf (a)} shows the PCA-transformed $z-$score data projected into the first two principal components, PC1-PC2 and classified according to the Leiden clustering performed in~\ref{sec:EC_annotation}. Panel {\bf (b)} shows the same projection colored with the final EC phenotypes according to our classification procedure, with : 1,320 Glycolytic ECs, 344 Stalk ECs, and 232 Tip ECs. 
    Axes include (in brackets) the percentage of covariance explained by each component.}
    \label{fig:A1}
\end{figure}


\subsection{RNA Velocity}
\label{subsec:rna-velo-pipeline_SM}

\subsubsection{RNA Velocity: a brief overview}
\label{subsubsec:rna-velocity-overview}

RNA velocity is a method to estimate the instantaneous rate of change in gene expression using unspliced and spliced scRNA-seq data from a static experiment~\cite{la2018rna,bergen2020generalizing}. For clarity, this method does not consist in following the same cell and record its state at two distinct times. Instead, it uses the unspliced RNA as a proxy for the recently produced transcript; and spliced RNA for the mature transcript pool. Formally, each gene $i$ follows a simple kinetic model:
\begin{align}
\emptyset \xrightarrow{\,\alpha_i(t)\,} u_i(t)
\xrightarrow{\,\beta_i\,} s_i(t)
\xrightarrow{\,\gamma_i\,} \emptyset\,,
\qquad
\left\{\begin{aligned}
    \frac{du_i}{dt} &= \alpha_i(t) - \beta_i u_i\,, \\
    \frac{ds_i}{dt} &= \beta_i u_i - \gamma_i s_i \,.
\end{aligned}\right.
\label{eq:gene-dynamics}
\end{align}
Here $u_i$ and $s_i$ represent the relative abundance of unspliced and spliced transcripts of the gene, and $t$ denotes a time coordinate that is internal to each cell. Importantly, in the RNA velocity fit algorithm, the time coordinate is treated as a parameter with no physical units. The conversion of $t$ from internal time units into physical time units will be discussed below in SM-\ref{subsubsec:rna-velocity-physical-time}.

In the steady state, and under a constant transcription rate $\alpha_i$, we obtain the relation $u_i^{\textrm{ss}}=(\gamma_i/\beta_i)s_i^{\textrm{ss}}$. This provides a useful baseline to determine when the spliced abundance is predicted to increase/decrease in a cell by studying the sign of $\beta_iu_i-\gamma_is_i$. However, this argument only works in the steady-state approach. In what follows, we use the so-called {\em dynamical mode} to fit the parameters of the model. This allows us to fit the data to transient trajectories in the $(u_i,s_i)$ phase space.

In the standard terminology~\cite{la2018rna}, RNA velocity is defined as the derivative of the spliced abundance
\begin{align}
    \nu_i:=\frac{ds_i}{dt}=\beta_iu_i-\gamma_is_i\,.\label{eq:classic-rna-velocity}
\end{align}
In our approach, for each cell $a$, the matrix entry $s_{ai}$ will inform the expression state of gene $i$, and the entry of $\nu_{ai}$ will give the RNA velocity vector. We therefore define expression using spliced abundance only and purposefully exclude unspliced (and ambiguous) counts. This choice follows from the standard formulation of RNA velocity and the operational outputs in \texttt{scVelo}, which is designed to estimate changing rates in spliced abundances, Eq.~\eqref{eq:classic-rna-velocity}. Nonetheless, unspliced counts remain essential inputs into the kinetic inference of $\nu_{ai}$ itself.

In what follows, we use the \texttt{dynamical} mode of \texttt{scVelo}~\cite{bergen2020generalizing}. This approach assumes two regimes of transcription for each gene termed {\em induction} and {\em repressed} phases. In the induction phase, the gene $i$ displays $\alpha_i=\alpha^{(0)}_i>0$. Once the gene is switched off at some internal time $t_i^*$ it enters the repressed phase with $\alpha_i=0$. Hence, we model the previous transcription function $\alpha_i(t)$ as a step-function involving parameters $\{\alpha^{(0)}_i,t^*_i\}$. The algorithm fits these two parameters together with the remaining gene-specific parameters, $\{\beta_i,\gamma_i\}$. In the fitting process, each cell in the dataset is treated as a sample. As a rule of thumb, we require that the number of samples (cells) is much larger than the number of parameters to fit. For more details, see~\cite{la2018rna,bergen2020generalizing,scVeloDocumentation}. Finally, RNA velocity  is then extrapolated for each cell by inputting the respective unspliced and spliced RNA abundances into~\eqref{eq:classic-rna-velocity}.

Notice that fitting for $\nu_{ai}$ returns a numerical derivative in relative abundance coordinates. However, this derivative is performed against an internal (unitless) time $t$. To convert between this internal time coordinate $t$ and physical time $\tphys$ (e.g., in units of hours) requires an external reference. In SM-\ref{subsubsec:rna-velocity-physical-time} we provide this mapping by using data on human mRNA half-life.

\subsubsection{RNA-velocity fit}
\label{sec:rna-velocity-fit}

\paragraph*{RNA Velocity Computation with \texttt{scVelo} --- } To compute RNA velocities for the population of ECs phenotypically labeled and retained above. We first generate spliced, ambiguous and unspliced transcript-count matrices from the original W-sample sequencing data. The BAM file produced from the W-sample alignment with \texttt{Cell Ranger} is processed using \texttt{velocyto}~\cite{la2018rna,velocytoWebsite}, which classifies aligned reads as spliced, unspliced and ambiguous, and stores the resulting count matrices in a \texttt{.loom} file. The W-sample \texttt{.loom} file contains $10{,}964$ cells and $36{,}601$ genes, together with the corresponding spliced, unspliced and ambiguous count layers.

\begin{figure}
    \centering
    \includegraphics[width=0.875\linewidth]{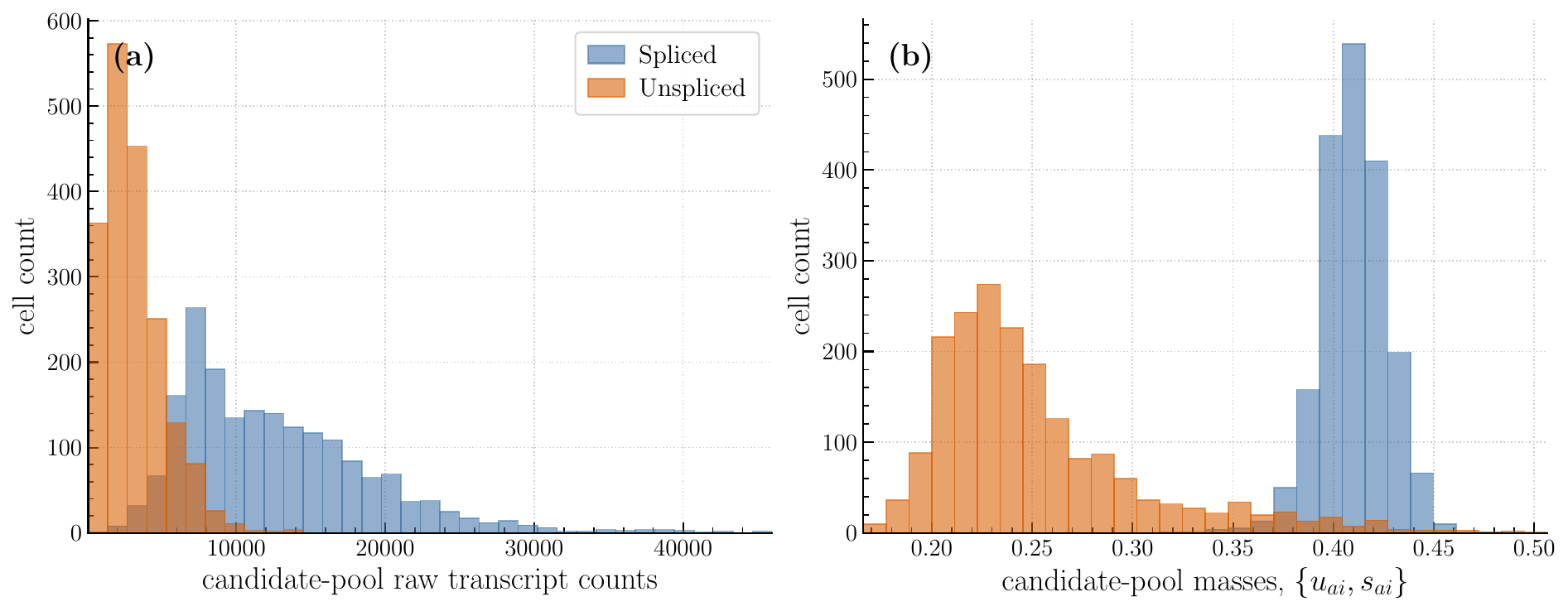}
    \caption{{\textbf{Transcript counts and densities.}} Panels {\bf (a-b)} are extracted from the \texttt{.loom} file. In panel {\bf (a)} we collect all the counts of unspliced and spliced transcripts across the $1896$ EC-annotated cells in the system and for all $36 601$ available genes and plot them in a histogram. In panel {\bf (b)} we normalized each entry by dividing by respective counts (unspliced and spliced) for all genes in a given cell, which returns de unspliced and spliced density masses, $u_{ai}=U_{ai}/\sum_j U_{aj}$ and $s_{ai}=S_{ai}/\sum_j S_{aj}$ thereby plotted in a histogram format.}
    \label{fig:A2}
\end{figure}

We next use the \texttt{scVelo} package~\cite{scVeloDocumentation} to compute RNA velocity using the W-condition \texttt{.loom} as the input. Cell identifiers present in the \texttt{.loom} file and in the phenotype-annotated ECs are converted to their common 10x cell-barcode representation and matched, yielding $1896$ uniquely matched ECs. Their respective phenotypic labels are retained for subsequent analysis. Note that such labels do not enter into the kinetic fitting of RNA velocities. For RNA-velocity estimation, we consider all $36 601$ genes available in the \texttt{.loom} file, i.e., we do not restrict the candidate pool of genes to the previously selected subset of $1940$ genes which were only used for phenotypic classification. Thus, the initial count matrices contain $1896$ cells and $36 601$ candidate features. Genes are then screened for spliced and unspliced counts, which are normalized following~\cite{la2018rna}. The final gene set used downstream is determined by quality criteria in the RNA-velocity fit, as described below.

The spliced and unspliced count matrices are separately normalized by dividing by the total spliced and total unspliced counts within each cell (see Fig.~\ref{fig:A2}). Genes are subsequently retained for kinetic fitting if they have at least 20 shared counts (i.e., between both spliced and unspliced RNA transcripts) in at least three cells and nonzero variance in both normalized layers. We do not impose any highly variable gene selection or fixed top-gene limit. After applying such filters, the normalization denominators are preserved. Then, we construct a neighborhood graph from the first 30 principal components of the \texttt{log1p}-transformed spliced densities, using \texttt{sc.pp.neighbors} with \texttt{n\_neighbors=30}. The resulting graph is used to obtain the row-normalized neighborhood operator through \texttt{scVelo}'s \texttt{get\_connectivities}, including self-contributions. We then compute neighborhood-averaged first moments of the linear spliced and unspliced densities and use these values for subsequent kinetic fitting. For clarity, this logarithmic transformation is only used for graph construction.

Gene-specific transcriptional dynamics are then inferred using \texttt{scv.tl.recover\_dynamics} with \texttt{use\_raw=False}. This procedure fits the dynamical model of transcription, splicing, and degradation (Eqs.~\eqref{eq:gene-dynamics}) to the observed spliced and unspliced mass densities and estimates the corresponding gene-specific kinetic parameters (see SM~\ref{subsubsec:rna-velocity-overview}). The velocities returned using \texttt{scv.tl.velocity} with \texttt{mode="dynamical"} are retained for diagnosis and comparison. In the downstream analysis, we evaluate $\nu_{ai}=\hat{\beta}_iu_{ia}-\hat{\gamma}_is_{ia}$ for each cell $a$ and each gene $i$, where we use the exported $\{\hat{\beta}_i,\hat{\gamma}_i\}$ from the \texttt{scVelo} algorithm (for the output of all kinetic parameters, see Fig.~\ref{fig:A3})\footnote{For completeness, the \texttt{scVelo} algorithm introduces a further fitting parameter, namely $c_i$ for each gene that rescales $u_{ai}\to \tilde{u}_{ai}=u_{ai}/c_i$. To return to the original equations~\eqref{eq:classic-rna-velocity}, which are written in the space of $(u_{ia},s_{ia})$, one needs to appropriately rescale the kinetic parameters with this fitted scaling factor. For simplicity, the reported values in Fig.~\ref{fig:A3} are already rescaled, thus corresponding directly to those that appear in Eqs.~\eqref{eq:classic-rna-velocity}.}.  
After applying the kinetic-fit quality and finite-value checks, which consist of discarding all genes with $R^2<0$ and \texttt{NaN} outputs in the estimated parameters, the expression and velocity matrices each contain $1896$ cells and $1873$ genes.

\begin{figure}
    \centering
    \includegraphics[width=0.92\linewidth]{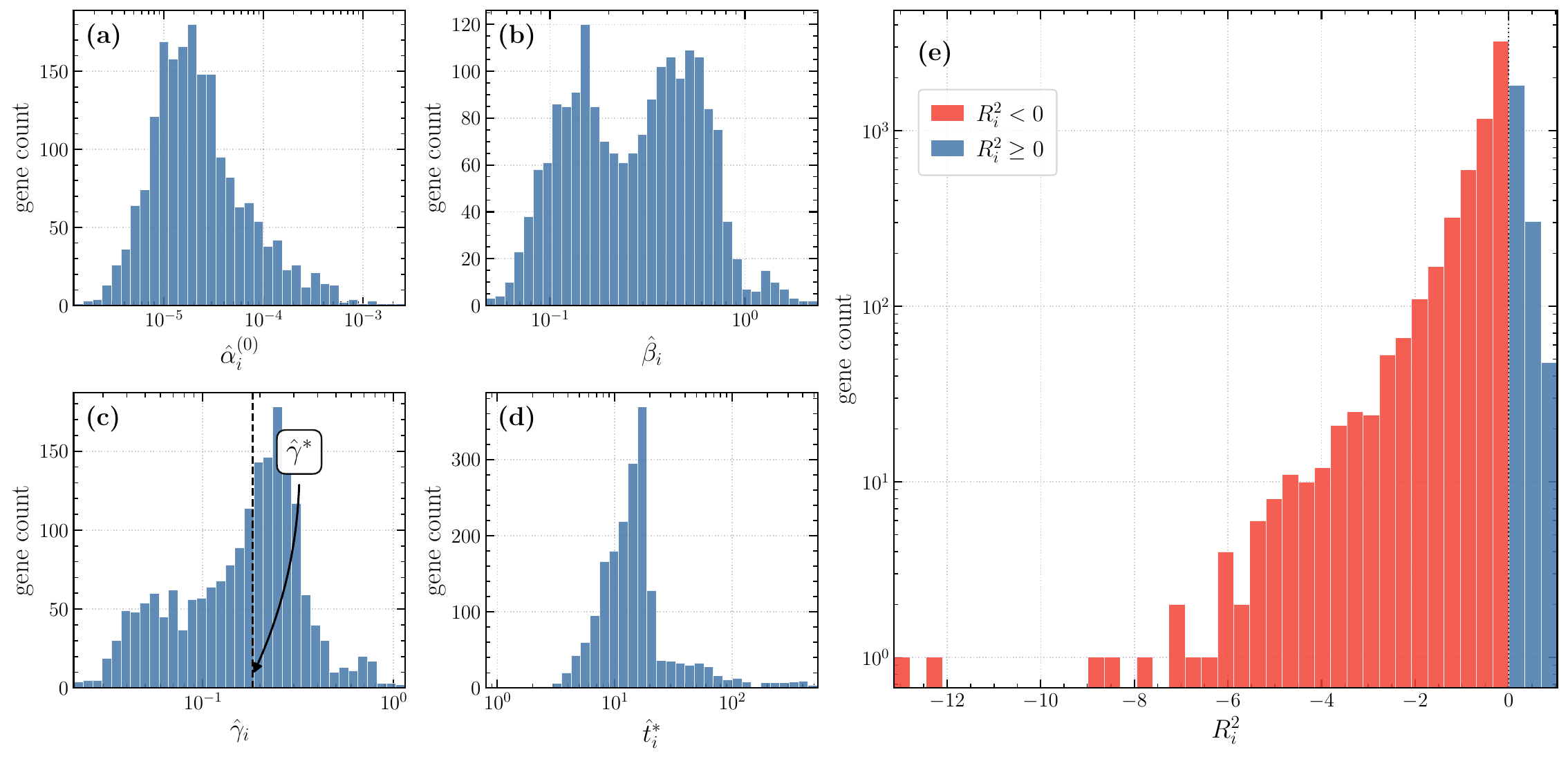}
    \caption{{\textbf{Kinetic parameters \& goodness of fit.}} Panels {\bf (a-d)} show the distributions of the gene parameters involved in the \texttt{scVelo} dynamical algorithm with logarithmically spaced axes for: transcription rate ($\alpha_i^{(0)}$), splicing rate ($\beta_i$), decay rate ($\gamma_i$) and internal time ($t_i^*$), respectively, for the $ 1873$ valid genes. The dashed vertical line in panel {\bf (c)} gives the typical decay rate, defined as $\hat{\gamma}^*=\operatorname{median}\{\gamma_i\}\simeq 0.183$ in units of inverse internal time. Panel {\bf (e)} shows the distribution of the statistic $R^2_i$ for all $8024$ preprocessed genes. Genes with $R_i^2\leq 0$ are colored red and are deemed invalid in the \texttt{scVelo} output, while bars with $R_i^2>0$ account for the valid genes. The dashed vertical line marks $R_i^2=0$.}
    \label{fig:A3}
\end{figure}

\paragraph*{Standardized expression \& velocity --- }

The final outputs of the \texttt{scVelo} pipeline outlined above correspond to two matrices or data-frames, namely $S=(s_{ai})$ and $V=(\nu_{ai})$, consisting of $1896$ cells and $1873$ genes each. We further split the sampled cells into a training subset and a test subset of $1516$ ($\sim 80\%$) and $380$ ($\sim 20\%$) cells, respectively. 

Since the training set of spliced density masses varies strongly across genes  (and so do their respective velocity estimates), we standardize the resulting data frames by centering all spliced density masses and normalizing by the standard deviation obtained from each column (gene), i.e.,
\begin{align}
    S\longmapsto Z=(z_{ai}) \quad \textrm{with}\quad z_{ai}=\frac{s_{ai}-\bar{s}_i}{\textrm{std}(s_i)}\,,\label{eq:normalized-expressions}
\end{align}
where $\bar{s}_i=\sum_b s_{bi}/M^{\circ}$ and $\operatorname{std}(s_i)^2=\sum_b (s_{bi}-\bar{s}_i)^2/(M^{\circ}-1)$, where $M^{\circ}=1516$ corresponds to the training subset. For velocities, we take
\begin{align}
    V\longmapsto W=(w_{ai}) \quad \textrm{with} \quad w_{ai}=\frac{\nu_{ai}}{\operatorname{std}(s_i)}\,,\label{eq:normalized-velocities}
\end{align}
i.e., without subtracting the velocity mean. Both $\bar{s}_i$ and $\operatorname{std}(s_i)$ are assumed to take fixed values. Hence, the relation $ds/dt=\nu$ implies $dz/dt=w$ in the new coordinate system. This assumption is consistent with viewing the sampled tissue is sufficiently representative of the genetic variation within ECs undergoing angiogenesis. The resulting matrices $Z,W\in\Reals^{M^{\circ}\times N}$, together with the associated EC phenotypic labels, provide the inputs for our subsequent Hopfield-network analysis. In Fig.~\ref{fig:A4}, we show the flow plots in expression-PC space for $Z$ and $W$.

\begin{figure}
    \centering
    \includegraphics[width=0.85\linewidth]{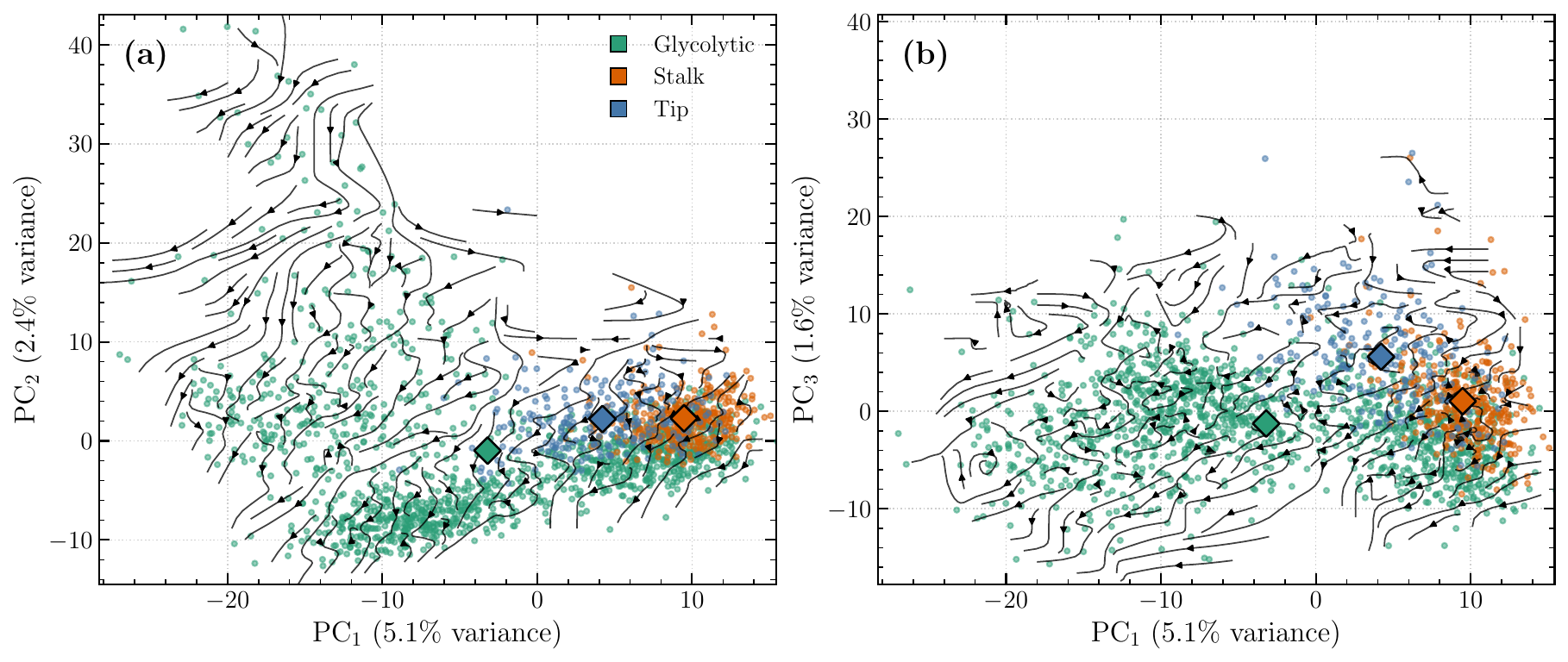}
    \caption{{\textbf{Principal components \& velocity flow representations.}} Panel {\bf (a)} shows the standardized spliced expression data projected into the first two principal components, PC$_1$-PC$_2$, together with a subset of transformed velocity field vectors captured by matrix $W$. Panel {\bf (b)} shows the equivalent projection for PC$_1$-PC$_3$. Each axis includes (in brackets) the percentage of covariance explained by each respective expression-PC. Centroids $(z_{\mathrm{G}},z_{\mathrm{S}},z_{\mathrm{T}})$ are too projected into PC space and depicted in corresponding colors.}
    \label{fig:A4}
\end{figure}


\subsubsection{From internal to physical time}
\label{subsubsec:rna-velocity-physical-time}

As discussed above, the dynamical fit from \texttt{scVelo} does not fix the time-scale of the system. Instead, the internal time $t$ used in Eqs.~\eqref{eq:classic-rna-velocity} has no usual physical units. We provide a rough calibration of these units by adopting a representative reference of mRNA half-life of approximately ten hours,  $h^*\simeq 10\,\textrm{hours}$, which was reported for human cell lines in~\cite{BNID104747}. This sets the physical timescale for the typical first-order degradation rate
\begin{align}
    \gamma^*_{\mathrm{ext}}=\frac{\ln 2}{h^*}\,.
    \label{eq:typipcal-half-life-to-typical-gamma}
\end{align}
From the output obtained via \texttt{scVelo}, we have a total of $1873$ values for $\hat{\gamma}_i$, one for each valid gene. We compute the typical value as $\hat{\gamma}^*=\textrm{median}(\{\hat{\gamma}_i\})$ and take
\begin{align}
    \lambda:=\frac{\gamma^*_{\mathrm{ext}}}{\hat{\gamma}^*}=\frac{\ln 2}{\hat{\gamma}^* h^*}\,.\label{eq:time-factor}
\end{align}
This result allows us to provide a rough relation between the internal (dimensionless) time $t$ used in the RNA-velocity fit and the physical time $\tphys$ (in hours):
\begin{align}
    t = \lambda \tphys\,.
\end{align}
From our results above (see Fig.~\ref{fig:A3}c), we estimate
\begin{align}
    \hat{\gamma}^*\simeq 0.18\,.
\end{align}
Using the value for $h^*$ reported above,
\begin{align}
    \gamma_{\mathrm{ext}}^{*} =\frac{\ln 2}{h^{*}}
      \simeq 0.07\,\mathrm{h}^{-1}\quad\textrm{and}\quad\lambda=\frac{\gamma_{\mathrm{ext}}^{*}}{\hat{\gamma}^{*}}
      \simeq 0.38\,\mathrm{h}^{-1}.
\end{align}
Hence, we obtain a rough mapping between the internal time $t$ used in the RNA velocity algorithm and the physical time $\tphys$ in units of hours, which is the relevant timescale for our biological system. In particular,
\begin{align}
    \tphys = \lambda^{-1}t    \simeq (2.64\,\mathrm{h})\,t \,.
\end{align}
Similarly, this transforms velocities as $W\mapsto W^{\textrm{phys}}=\lambda W$.

\subsubsection{Calibration of the Glauber timescale $\tau$}
\label{subsubec:calibration_glauber_tau}

The Glauber timescale $\tau$, introduced in Sec.~\ref{subsec:sm_glauber} below, sets the microscopic timescale of spin-flipping that characterizes the dynamical Hopfield network that we use to model our system. In this section, we calibrate this parameter using the relaxation timescales of the estimated kinetic parameters obtained in the RNA velocity fit. Consider a fixed gene $i$ and recall the RNA velocity model provided by Eqs.~\eqref{eq:classic-rna-velocity}. Here, a small perturbation from the steady-state applied to the unspliced and spliced population densities $u\to u+\delta u$ and $s\to s+\delta s$ (occurring at initial internal time $t=0$) evolves as
\begin{align}
\frac{d}{dt}\begin{pmatrix}
\delta u_i\\
\delta s_i
\end{pmatrix}=\begin{pmatrix}
-{\beta}_i & 0\\
{\beta}_i & -{\gamma}_i
\end{pmatrix}\begin{pmatrix}
\delta u_i\\
\delta s_i
\end{pmatrix}\,.
\end{align}
The relaxation eigenvalues are directly read from the system above and are given by $-\beta_i$ and $-\hat{\gamma}_i$. In particular, for the spliced response (which sets the basis of our expression matrix $Z$), 
\begin{align}
    \delta s_i (t) = A_i e^{-\beta_i t}+B_ie^{-\gamma_i t}
\end{align}
where the amplitudes depend on the initial condition. Hence, the slowest available relaxation mode has an internal timescale given by
\begin{align}
    \theta_i= \frac{1}{\min({\beta}_i,{\gamma}_i)}.
\end{align}
This mode controls the asymptotic relaxation (as long as its associated amplitude is nonzero). We assign the physical Glauber timescale to the median of these gene-specific times computed over the $N=1873$ retained genes in the dataset and divide by the scaling factor $\lambda$ to return $\tau$ in units of hours. That is,
\begin{align}
    \tau = \operatorname{median}\{\hat{\theta}_i/\lambda\}=\operatorname{median}\left\{\frac{1}{\lambda \min(\hat{\beta}_i,\hat{\gamma}_i)}\right\}\simeq 15.0\,\textrm{hours}\,.
\end{align}
Matching $\tau$ to the typical slow RNA relaxation provides a physically motivated calibration of the effective spin dynamics. However, we note that this correspondence is a modeling assumption. In particular, this assumption does not establish that thresholded (or signed) gene expression follows Glauber dynamics. Moreover, as we will see in the remainder of the paper, interactions in the resulting Hopfield network can produce collective relaxation times and oscillation periods different from the predicted $\tau$.

\section{Dynamics of a nonreciprocal Hopfield net}
\label{sec:nonrep_hopfield_sm}

In this section, we provide the technical details of the nonreciprocal Hopfield-net approach introduced in Sec.~\ref{sec:nonrep_hopfield} from the main text. We first formulate the asynchronous Glauber dynamics as a continuous-time Markov chain and discuss how nonreciprocal couplings can break detailed balance. We introduce the Mattis gage and the empirical Gram matrix, use them to organize genes into four relative-sign groups, and derive the dynamical mean-field equations for the phenotype overlaps. We briefly discuss the local stability of the zero-overlap solution.

\vspace{0.5em}
\subsection{Glauber dynamics}
\label{subsec:sm_glauber}

We implement the stochastic dynamics of the spin-like states $\bS\in\{-1,+1\}^N$ as a continuous-time Markov chain (CTMC) with asynchronous single-spin flips. We use the local field acting on the spins given by Eq.~\eqref{eq:h} in the main text and introduce spin-flip transitions,  $S_i\to -S_i$, at rates
\begin{align}
    w_i(\bS)=\frac{1}{2\tau}\left[1-S_i\tanh\left(h_i(\bS)\right)\right]\quad \textrm{for each }i\in\{1,\ldots,N\}\,,\label{eq:transition-rates}
\end{align}
where $\tau$ sets the microscopic update timescale. 
Note that in~\eqref{eq:transition-rates} we absorbed the usual inverse-temperature factor into the field, or, equivalently, into the overall scale of the mixing matrix $\mix$ introduced in Eq.~\eqref{eq:JHop} of the main text. Note that we use the convention of $J_{ii}=0$ throughout, so that $h_i(\bS)\sum_{j\neq i}J_{ij}S_j$, which is independent of $S_i$.

Denoting by $P_t(\bS)$ the probability of observing state $\bS$ at time $t$, the master equation reads
\begin{align}
    \frac{d}{dt}P_t(\bS)=\sum_{i}\Big[&w_i\left(\F_i(\bS)\right)P_t\left(\F_i(\bS)\right) -w_i(\bS)P_t(\bS)\Big]\,,\label{eq:MC}
\end{align}
where we introduced $\F_i(\bS)=(S_1,\ldots,-S_i,\ldots,S_N)$ as the spin-flip operator for the $i-$th spin in $\bS$, and the transition rates $w_i$ are read from~\eqref{eq:transition-rates}.

For a reciprocal network, $J=J^{\intercal}$, the system admits an energy representation with Hamiltonian
\begin{align}
    H(\bS):=-\frac{1}{2}\bS^{\intercal}J\bS\,,
\end{align}
where the Hamiltonian satisfies $H(\F_i(\bS))-H(\bS)=2S_i h_i(\bS)$. Then rates in
Eq.~\eqref{eq:transition-rates}  obey
\begin{align}
    \frac{w_i(\bS)}{w_i(\F_i(\bS))}
    =\exp\!\left[-H(\F_i(\bS))+H(\bS)\right]\,,
\end{align}
and, consequently, the Boltzmann distribution $\pi(\bS)\propto e^{-H(\bS)}$
satisfies the detailed-balance condition,
\begin{align}
    w_i(\bS)\pi(\bS)
    =w_i(\F_i(\bS))\pi(\F_i(\bS))\,.
    \label{eq:DB}
\end{align}
However, for a generic asymmetric $J$, no general scalar energy function of the original spins generates
the dynamics, and the condition~\eqref{eq:DB} need not hold. Hence, for nonreciprocal $J$, stationary distributions
can sustain nonzero edge currents
\begin{align}
    \mathcal{J}_i(\bS)
    :=w_i(\bS)P_{\mathrm{ss}}(\bS)
      -w_i(\F_i(\bS))P_{\mathrm{ss}}(\F_i(\bS))\,.
    \label{eq:configuration-current}
\end{align}
In this paper, we relax the detailed balance condition by allowing asymmetric interaction matrices $J$, and thus edge currents are possible in our set-up.


\subsection{Dynamical mean-field approach}
\label{subsec:dmft_sm}

\subsubsection{Empirical Gram matrix \& Mattis transform}
\label{subsubsec:mattis}

Let $\bXi:=\left(\begin{matrix}
    \bxip & \bxis & \bxit \end{matrix}\right)
\in\{-1,+1\}^{N\times |\Gamma|}$, which collects the three empirical memory patterns obtained from Eq.~\eqref{eq:centroid-to-memory-states} in the main text. Their empirical
Gram (or correlation) matrix is defined as
\begin{align}
    \mathcal{Q}:=\frac{1}{N}\bXi^{\intercal}\bXi\in\Reals^{|\Gamma|\times|\Gamma|}\,,\quad \textrm{with entries}\quad
    \mathcal{Q}_{\mu\nu}
    =\frac{1}{N}{\bximu}^{\intercal}\bxinu\,.
    \label{eq:Gram-matrix-sm}
\end{align}
This matrix is symmetric, has a unit diagonal, and when empirical memory patterns are non-orthogonal, off-diagonal terms in $\mathcal{Q}$ are nonzero.  We note that, using this compact notation, the interaction matrix introduced in Eq.~\eqref{eq:JHop} in the main text is written as\footnote{To be consistent, one needs to first take $\tilde{J}=\bXi\mix\bXi^{\top}/N$ and then set $J=\tilde{J}-\operatorname{diag}(\tilde{J})$ so that we set $J_{ii}=0$ appropriately.} $J=\bXi \mix \bXi^{\top} /N$.

The Mattis transformation~\cite{mattis1976solvable} is a common site-wise spin relabeling used in the context of Hopfield nets and may be regarded as a gauge fix. Let us introduce $D_{\mathrm{G}}\in\{0,\pm1\}^{N\times N}$ whose entries are defined as $(D_{\mathrm{G}})_{ii}=\xi^{\mathrm{(G)}}_i$ and zero for all $i\neq j$. Then, without loss of generality, we transform the states, memory patterns, interaction matrix and local field consistently as\footnote{We use the Hadamard product notation, $\circ$, to generally denote $(x\circ y)_i= x_iy_i$.}
\begin{align}
     \bS&\mapsto \bS'=D_{\mathrm{G}}\bS=\bxip\circ\bS\,,\nonumber\\
     \bximu&\mapsto {\bximu}'=D_{\mathrm{G}}\bximu=\bxip\circ\bximu\,,\nonumber\\
     J&\mapsto J'=D_{\mathrm{G}}JD_{\mathrm{G}}\,,\nonumber\\
     \bh&\mapsto \bh'=D_{\mathrm{G}}\bh\,.
    \label{eq:mattis-1}
\end{align}
Note that $D_{\mathrm{G}}^{2}=I_N$, which means that the glycolytic pattern is mapped to
${\bxip}'=\1_N$, while the other two patterns become 
${\bxis}'=\bxip\circ\bxis$ and ${\bxit}'=\bxip\circ\bxit$. Moreover, the
interaction matrix retains the same form,
\begin{align}
    J'=\frac{1}{N}\bXi'\mix{\bXi'}^{\intercal}\,,
\end{align}
so the mixing matrix $\mix$ is invariant under the Mattis gauge choice. Because
$S_i'h_i'=S_i h_i$, the Glauber rates and, hence, the stochastic dynamics are also
unchanged.

On the other hand, the overlaps defined in Eq.~\eqref{eq:overlap-def} of the main text and the Gram matrix~\eqref{eq:Gram-matrix-sm} are also invariant:
\begin{align}
    m_\mu'(\bS')
    &=\frac{1}{N}({\bximu}')^{\intercal}\bS'
      =m_\mu(\bS)\,,\nonumber\\
    \mathcal{Q}_{\mu\nu}'
    &=\frac{1}{N}({\bximu}')^{\intercal}{\bxinu}'
      =\mathcal{Q}_{\mu\nu}\,.
    \label{eq:mattis-invariants}
\end{align}

From this point onward, we only work in the Mattis transformed spin space, and, for simplicity, we drop the primes for the rest of this section and the whole of the main text. 

We note that, under this choice of gauge, the overlap with the glycolytic state becomes a global magnetization: $\mP=(1/N)\1_N^{\intercal}\bS\equiv m$. On the other hand, the off-diagonal Gram entries 
read as
\begin{align}
    \mathcal{Q}_{\textrm{GS}}&=\frac{1}{N}\1_N^{\intercal}\bxis\,,\qquad
    \mathcal{Q}_{\textrm{GT}}=\frac{1}{N}\1_N^{\intercal}\bxit\,,\qquad
    \mathcal{Q}_{\textrm{ST}}=\frac{1}{N}{\bxis}^{\intercal}\bxit\,.
    \label{eq:transformed_Gram}
\end{align}
It is useful to collect these quantities into a vector
$\bQ:=(1,\mathcal{Q}_{\textrm{GS}},\mathcal{Q}_{\textrm{GT}},
\mathcal{Q}_{\textrm{ST}})^{\intercal}$. For clarity, we note that the Mattis gauge does not remove a
phenotypic memory: all three memory patterns and all three overlaps remain. Instead, its main
advantage is that, in our system of $|\Gamma|=3$ memory states, fixing $\bxip=\1_N$ reduces the possible relative gene signatures of the form $(\xi_i^{\mathrm{(G)}},\xi_i^{\mathrm{(S)}},\xi_i^{\mathrm{(T)}})$ from $2^3=8$ down to $2^2=4$ possibilities.

\subsubsection{Empirical correlations between memories \& group overlaps}
\label{subsubsec:empirical-corrs}

Geometrically, a non-diagonal Gram matrix $\mathcal{Q}$ indicates that
the empirical memory patterns are not mutually orthogonal in the
$N$-dimensional state space. Then, the overlap components in $\bm$ (as defined in Sec.~\ref{subsubsec:set-up} of the main text) are
projections onto a nonorthogonal set of memory directions and cannot, in general, be interpreted as independent axes. To see
this explicitly, let $\boldsymbol{e}_\nu\in\mathbb{R}^{3}$ denote the
canonical basis vector associated with memory $\nu$, such that
$\bxinu=\bXi\boldsymbol{e}_\nu$. Using
$\bm(\bS)=(1/N)\Xi^{\intercal}\bS$, we then obtain
\begin{align}
    \bm(\bxinu)= \frac{1}{N}\bXi^{\intercal}\bxinu =
    \frac{1}{N}\bXi^{\intercal}\bXi\boldsymbol{e}_\nu =
    \mathcal{Q}\boldsymbol{e}_\nu\,.\label{eq:memory-as-Gram-column}
\end{align}
Therefore, each memory state is represented in overlap space by the
corresponding column of $\mathcal{Q}$. A spin state that coincides with one of the memories generally also has nonzero overlaps with the others. There exist methods to compensate for the correlations among stored patterns, for example via the pseudoinverse prescription (see~\cite{hertz1991introduction} pp. 49-53). However, we purposefully retain the nonorthogonal representation induced by the correlations of the Gram matrix so that  the overlaps correspond directly to projections onto the observed phenotypic signatures.

We also note that the presence of memory correlations is not an additional source of nonreciprocity. To see this, consider our definition of the interaction matrix $J$ as given in Eq.~\eqref{eq:JHop} in the main text. Using the compact notation introduced above,
\begin{align}
    J-J^{\intercal}
    =\frac{1}{N}\bXi\left(\mix-\mix^{\intercal}\right)\bXi^{\intercal}\,.
    \label{eq:J-asymmetry-source}
\end{align}
Therefore $\mix=\mix^{\intercal}$ implies $J=J^{\intercal}$ regardless
of the off-diagonal entries of $\mathcal{Q}$. From a dynamical viewpoint, correlations between the memory states do not
independently break detailed balance.

To incorporate the empirical correlations into the dynamical mean-field reduction, it is useful to consider the pairs $(\xis_i,\xit_i)\in\{-1,+1\}^{2}$ for each gene label $i$. The four possible relative signatures of these pairs are classified as
\begin{align}
    A: (+,+)\,;\;\;B: (-,+)\,;\;\;C:(+,-)\,\;\;\textrm{and}\;\;D:(-,-)\,.\label{eq:group_classification}
\end{align}
Denote by $\rho_X=N_X/N$ the respective frequencies of occurrence of pairs $X\in\{A,B,C,D\}$ as obtained directly from the data. Then, their relation to the
empirical Gram matrix and, in particular, to the vector $\bq$ defined above is fully derived in Box~\hyperref[box:2]{2}. This useful relation is quoted in the main text within Sec.~\ref{subsec:dmft}.

Given a microstate $\bS$, every entry $S_i$  belongs to a single group $X\in\{A,B,C,D\}$. This allows us to define {\em group overlaps} as operators on the state $\bS$, which we denote as
\begin{align}
    g_X:=\frac{1}{N}\sum_{i\in X} S_i=\rho_X \left(\frac{1}{N_X}\sum_{i\in X}S_i\right) =\rho_X\overline{S_X}\,.\label{eq:def-m_X}
\end{align}
Plugging~\eqref{eq:def-m_X} into the definition of the overlaps (see Eq.~\eqref{eq:overlap-def} and Sec.~\ref{subsubsec:set-up} of the main text), 
\begin{align}
    \mP&= m=\frac{1}{N}\sum_X\sum_{i\in X} S_i = \sum_X q_X\\\nonumber
    &=g_A+g_B+g_C+g_D\,,
\end{align}
while for the other two phenotype overlaps,
\begin{align}
    \mS&=g_A-g_B+g_C-g_D\,\quad \textrm{and}\\
    \mT&=g_A+g_B-g_C-g_D\,,
\end{align}
where we used the group classification~\eqref{eq:group_classification}.
Thus, for $\bg:=(g_A,g_B,g_C,g_D)^{\intercal}$ and
$\bm:=(\mP,\mS,\mT)^{\intercal}$,
\begin{align}
    \bm = \mathfrak{H}_{3\times 4}\bg\,,\quad \textrm{with }\; \mathfrak{H}_{3\times 4}:=\left(\begin{matrix} 1 & 1 & 1 & 1 \\ 1 & - 1 & 1 & -1 \\ 1 & 1 & -1 & -1\end{matrix}\right)\,.\label{eq:q-to-m}
\end{align}
The rectangular matrix $\mathfrak{H}_{3\times 4}$ is the top $3 \times 4$ block of $H_4$. We also note the useful identity $\mathfrak{H}_{3\times 4}\mathfrak{H}_{3\times 4}^{\top} = 4 I_3$. For simplicity of notation, in the main text, we denote $\mathfrak{H}\equiv \mathfrak{H}_{3\times 4}$.


\vspace{2em}
\noindent
\begin{mytextbox}{Empirical group frequencies \& relation with the transformed Gram matrix}
\phantomsection
\label{box:2}

Here, we derive the relation between $\brho$ and $\bq$ reported in Sec.~\ref{subsec:dmft} from the main text.

Under the Mattis gauge, the off-diagonal elements of the Gram matrix $\mathcal{Q}$ correspond to the moments of the transformed memory entries:
\begin{align}
    \mathcal{Q}_{\textrm{GS}}\to \overline{ \xis}\,,\;\;\mathcal{Q}_{\textrm{GT}}\to \overline{\xit}\,\,\textrm{and}\;\; \mathcal{Q}_{\textrm{ST}}\to \overline{\xis \xit}\,,
\end{align}
where the overline notation $\overline{\,\cdot\,}$ denotes an empirical average over genes (not over stochastic trajectories).

From the pair group classification introduced in~\eqref{eq:group_classification}, we note that $\xis=+$ in groups $A$ and $C$, and $\xis=-$ in $B$ and $D$; $\xit=+$ in groups $A$ and $B$, and $\xit=-$ in $C$ and $D$; and $\xis\xit=+$ in groups $A$ and $D$, and $\xis\xit=-$ in $B$ and $C$; which gives
\begin{align}
    \mathcal{Q}_{\textrm{GS}}=(\rho_A+\rho_C)-(\rho_B+\rho_D)\,,\quad
    \mathcal{Q}_{\textrm{GT}}=(\rho_A+\rho_B)-(\rho_C+\rho_D)\,,\quad\textrm{and}\quad
    \mathcal{Q}_{\textrm{ST}}=(\rho_A+\rho_D)-(\rho_B+\rho_C)\,.\nonumber
\end{align}
Together with the normalization condition, $\sum_{X}\rho_X=1$, we arrive at:
\begin{align}
   \left( \begin{matrix}1\\\mathcal{Q}_{\textrm{GS}}\\\mathcal{Q}_{\textrm{GT}}\\\mathcal{Q}_{\textrm{ST}}\end{matrix} \right) = \left(\begin{matrix} 1 & 1 & 1 & 1 \\ 1 & -1 & 1 & -1 \\ 1 & 1 & -1 & -1 \\ 1 & -1 & -1 & 1 \end{matrix}\right)\left(\begin{matrix}\rho_A\\\rho_B\\\rho_C\\\rho_D\end{matrix}\right)\equiv H_4\brho \Longleftrightarrow \brho =\frac{1}{4}H_4\bQ\,,
\end{align}
where we used $H_4$ as the Hadamard-Walsh matrix of order four, which here transforms from vector $\brho$ to the vector $\bQ$, 
and obeys $H_4^2=4I_4$ (and, hence, $H_4^{-1}=\frac{1}{4}H_4$).

\end{mytextbox}
\vspace{1em}


The fourth component that would result from extending~\eqref{eq:q-to-m} using the $H_4$ instead is written as
$\tilde m:=g_A-g_B-g_C+g_D$. However, $\tilde{m}$ does not enter the interaction field and, hence, it does not feed back into the three-dimensional dynamical mean-field that governs $\bm(t)$.
We also note that physical overlap vectors do not generally fill the entire cube
$[-1,1]^3$. Since $|g_X|\leq\rho_X$ (which follows from~\eqref{eq:def-m_X}), their convex domain is
\begin{align}
    \mathcal{D}
    :=\left\{
        \mathfrak{H}_{3\times4}\bg:\ |g_X|\leq\rho_X
        \text{ for every }X
    \right\}
    \subseteq[-1,1]^3\,.
    \label{eq:physical-overlap-domain}
\end{align}
Initial conditions used in the DMFT approach are intended to represent cellular states, and should therefore be chosen inside $\mathcal{D}$.

\subsubsection{DMFT by group overlaps}
\label{subsubsec:dmft_groups}

We now return to working with dynamical ensemble averages by defining
\begin{align}
    g_X(t):=\left\langle g_X(\bS(t))\right\rangle\,,\qquad
    \bm(t):=\left\langle\bm(\bS(t))\right\rangle\,.
    \label{eq:averaged-overlaps}
\end{align}
Here, brackets denote averages over realizations of the stochastic trajectories that result from the Glauber process. Note that linearity preserves the relation 
$\bm(t)=\mathfrak{H}_{3\times4}\bg(t)$.


\vspace{2em}
\noindent
\begin{mytextbox}{Group local fields from group overlaps}
\phantomsection
\label{box:3}

Here we derive the relation between the group fields $\{h_X\}$ and the overlaps.

Using the definitions of interaction matrix and local field, Eqs.~\eqref{eq:JHop}-\eqref{eq:h} from the main text, we have that
\begin{align}
    h_i  = \sum_{\mu,\nu}\xi^{(\mu)}_i\mix_{\mu\nu}m_{\nu}=\bXi_i^{\intercal}\mix\bm\,,
\end{align}
where we recall that, under the Mattis gauge, $\bXi_i^{\intercal}:=\left(\begin{matrix} 1 & \xi_i^{(\textrm{S})} & \xi_i^{(\textrm{T})}
\end{matrix}\right)$. 
Next, we classify indices $i$ as belonging to the groups $X\in\{A,B,C,D\}$ defined in~\eqref{eq:group_classification}. To do so, we introduce the sign variables $(s_{\textrm{S}|X},s_{\textrm{T}|X})$ such that
\begin{align}
   (s_{\textrm{S}|A},s_{\textrm{T}|A})=(+,+);\quad   (s_{\textrm{S}|B},s_{\textrm{T}|B})=(-,+); \quad  (s_{\textrm{S}|C},s_{\textrm{T}|C})=(+,-)\quad  (s_{\textrm{S}|D},s_{\textrm{T}|D})=(-,-)\,.
\end{align}
We define $\bs_X^{\intercal}:=\left(\begin{matrix}1 & s_{\textrm{S}|X} &  s_{\textrm{T}|X}\end{matrix}\right)$, which allows us to classify the local field as perceived by each group in $X$,
\begin{align}
    h_X:=h_{i\in X} =\bs_X^{\intercal}\mix \bm\,.
\end{align}
Stacking these four fields in $\tilde{\bh}^{\intercal}:=\left(\begin{matrix} h_A & h_B  & h_C & h_D\end{matrix}\right)$, we obtain
\begin{align}
    \tilde{\bh}=\mathfrak{H}_{3\times 4}^{\intercal}\mix \bm,\quad \;\textrm{ with }\quad\; \mathfrak{H}_{3\times 4}^{\intercal}=\left(\begin{matrix} \bs_A^{\intercal}  \\ \bs_B^{\intercal}\\ \bs_C^{\intercal}\\ \bs_D^{\intercal}\end{matrix}\right) = \left(\begin{matrix}1 & 1 & 1 \\ 1 & -1 & 1 \\ 1 & 1 & -1 \\ 1 & -1 & -1 \end{matrix}\right).\label{eq:group-field-from-nonrepJ}
\end{align}

\end{mytextbox}


Multiplying the master equation~\eqref{eq:MC} by $S_i$ and summing over all
configurations gives the first-moment equation
\begin{align}
    \frac{d}{dt}\langle S_i\rangle = -\frac{1}{\tau}\langle S_i\rangle+\frac{1}{\tau}\langle \tanh\left( h_i\right)\rangle.\label{eq:mean-spin-evol}
\end{align}

We separately consider the pair groups classified into $X\in \{A,B,C,D\}$ and introduce the mean field assumption over groups: $\langle \tanh(h_{i\in X})\rangle\approx \tanh({h}_X)$, where ${h}_X$ denotes the mean field as perceived by the spins in the group $X$ (see Box~\hyperref[box:3]{3}). 
Summing over indices $i\in X$ on each side of Eq.~\eqref{eq:mean-spin-evol} and dividing by the total number of spins,
\begin{align}
    \frac{d}{dt}g_X  \approx - \frac{1}{\tau}g_X+\frac{\rho_X}{\tau}\tanh({h}_X)\,,\label{eq:DMFT-q-overlaps-vs-group-fields}
\end{align}
where we used definition~\eqref{eq:def-m_X}. 

Applying $\mathfrak{H}_{3\times4}$ to
Eq.~\eqref{eq:DMFT-q-overlaps-vs-group-fields} from the left, recalling~\eqref{eq:q-to-m} and substituting~\eqref{eq:group-field-from-nonrepJ}, we obtain the closed phenotype-level
dynamics:
\begin{align}
    \tau\frac{d}{dt}\bm \approx -\bm+ \mathfrak{H}_{3\times 4}\left(\brho\circ \tanh(\mathfrak{H}_{3\times 4}^{\intercal}\mix \bm)\right)\,.\label{eq:DMFT-m-overlaps-vs-group-fields}   
\end{align}
Above, $\tanh(\cdot)$ acts component-wise and $\circ$ denotes the Hadamard product. Within the mean-field approximation, the above result may be written as an equality, as it is done in the main text. 

\subsubsection{Stability of the trivial solution}
\label{subsubsec:trivial_solution}

We end this section by noting that the system in Eq.~\eqref{eq:DMFT-m-overlaps-vs-group-fields-main-text} from the main text always admits the zero-overlap solution, $\bm_0^*=\0$, which we term the trivial solution. Letting $\bm=\bm^*_0+\delta\bm$, linearization around the trivial point gives
\begin{align}
\tau\frac{d}{dt}\delta\bm
    =\left(-I_3+\mathcal{K}\right)\delta\bm    +O\left(\|\delta\bm\|^3\right)\,,
    \quad \textrm{where}\quad
\mathcal{K}:=\mathfrak{H}_{3\times4}\operatorname{diag}(\brho)
       \mathfrak{H}_{3\times4}^{\intercal}\mix\,.
    \label{eq:linearized-DMFT}
\end{align}
After some algebra, it is possible to show that $\mathfrak{H}_{3\times4}\operatorname{diag}(\brho)
\mathfrak{H}_{3\times4}^{\intercal}=\mathcal{Q}$, which gives the compact linearized map via $\mathcal{K}=\mathcal{Q}\mix$.
This solution is locally stable for 
\begin{align}
\max_{\alpha}\textrm{Re}\left(\lambda_{\alpha}(\mathcal{K})\right)<1\,,\label{eq:unstability-condition}
\end{align}
where $\{\lambda_{\alpha}(\mathcal{K})\}$ are the eigenvalues of the matrix $\mathcal{K}$. 
The eigensystem of $\mathcal{K}$ is an important object of study. For example, a real leading eigenvalue corresponds to a stationary linear instability, whereas a complex-conjugate pair crossing the boundary may lead to a Hopf bifurcation when the necessary transversality and
nondegeneracy conditions hold (see~\cite{xue2025critical} for a thorough discussion of a closely related example with $|\Gamma|=2$). 

Importantly, this criterion delineates the regions of parameter space in which the trivial solution is locally stable or unstable, but it does not determine the full nonlinear phase diagram. Moreover, the empirically inferred mixing matrix $\hat{\mix}$ locates the fitted model to a single point in parameter space. However, in order to quantify the proximity to criticality, one requires additional control of the system and exploration of the parameter space. Finally, in Sec.~\ref{subsec:model-outputs}, we discuss how to search for nontrivial fixed points (once the optimal mixing matrix is found) and study their stability and dynamics following a numerical search.

\subsection{Validation and statistical diagnostics of our fit}
\label{subsec:SM-mixing-matrix-fit}

In this section, we discuss the statistical methods and diagnostics applied to the panel of $N=1873$ genes and the training and test partitions of $1516$ and $380$ ECs, respectively denoted by $\train$ and $\test$. The dynamics, both DMFT and Glauber, use the calibrated $\tau\simeq 15.0$ hours. Mixing matrices are inferred for $\hor$ values between 2 and 8 hours, which  is biologically motivated, as discussed in the main text under Sec.~\ref{subsec:mix_matrix_fit}. Here onward, we express all time variables in physical units of hours.

\paragraph*{Horizon and target diagnostics ---}
Recall that the overlap targets are constructed by thresholding the extrapolation of expressions given by $Z+\lambda\hor W$, then using standardization, applying the sign transformation from Eq.~\eqref{eq:Z-to-spins} in the main text, and finally using the Mattis transformation with the same memory references as in the initial states. This yields two sets of overlap vectors $\{\hat{\bm}_a^{0}\}$ for the initial and $\{\hat{\bm}^{\hor}_a\}$ for the final overlaps for either $a\in\train$ or $a\in\test$.

A numerical constraint can be obtained directly from the DMF Eqs.~\eqref{eq:DMFT-m-overlaps-vs-group-fields}. Let us rewrite these equations component-wise by renaming the nonlinear $\tanh$ contribution as $\boldsymbol{F}(\bm)$, such that,
\begin{align}
    \tau \frac{d}{dt} m_{\mu} = -m_{\mu}+ F_{\mu}(\bm)\,,
\end{align}
then, the boundedness of the $\tanh$ implies that $|F_\mu|\leq1$. On the other hand, integration gives
\begin{align}
m_\mu(\hor) =e^{-\hor/\tau}m_\mu(0) +\frac{1}{\tau}\int_0^\hor e^{-(\hor-t)/\tau}F_\mu(\bm(t))dt\,.
\end{align}
Consequently, every DMF-predicted endpoint satisfies
\begin{align}
\left|m_\mu(\hor)-e^{-\hor/\tau}m_\mu(0)\right| \leq 1-e^{-\hor/\tau}\,.\label{eq:dmft-validity-test}
\end{align}
For each cell $a$ in $\train$ and each selected horizon value $\hor$, we substitute the per-cell values, $m(\hor) \to {m}_a^{\hor}$ and $m(0) \to {m}_a^{0}$, on the left hand side of~\eqref{eq:dmft-validity-test} (and for all $\mu\in\{\mathrm{G},\mathrm{S},\mathrm{T}\}$) and check whether the upper bound is respected. We compute the fraction of cells whose target violates this bound in any of its components (using an absolute tolerance of $10^{-12}$) and checked that \textit{all} cells comply with the bound for all values of $\hor>2$ hours.

We also report that the percentage of spins that flip in going from $S(0)=\operatorname{sign}(Z)\to \bS(\hor)=\operatorname{sign}(Z+\lambda \hor W)$ ranges from $5\%$ for $\hor=2$ hours up to $20\%$ for $\hor = 8$ hours.

\paragraph*{Optimization and accuracy ---}
For each horizon value $\hor$, we minimize the objective function
\begin{align}
\mathcal L_{\hor}(\mix)=\sum_{a\in\mathcal T}g_a
\lVert
\Phi_{\mix}^{\hor}(\hat{\bm}_a^0)-\hat{\bm}_a^{\hor}
\rVert^2\,,\quad\textrm{with}\quad 
g_a=\frac{|\mathcal T|}{3n_{\mu(a)}}.
\label{eq:min-squares}
\end{align}
where $\mu(a)\in\{\mathrm{G},\mathrm{S},\mathrm{T}\}$ denotes the label of the phenotype of cell $a\in\train$, and $n_{\mu(a)}$ is the total number of training-cell counts with $\mu(a)$ as its phenotype. In this way, the pre-factor $g_a$ equilibrates the participation of each phenotype in the dataset. In other words, the weights given by $\{g_a\}$ provide the three phenotypes with equal total contributions to the objective. Notice $g_a$ is preserved across horizons and is model-invariant.

We carried out this optimization problem using \texttt{scipy.optimize.least\_squares}, with \texttt{x\_scale="jac"}, termination tolerance $10^{-8}$ and \texttt{max\_nfev=800}. Four distinct random initial conditions are used in total for each model and horizon. The nonlinear DMFT flow is integrated using Runge--Kutta (RK4), first with time-steps of $0.05\tau$. Then, for every fitted matrix, we recompute the held-out endpoints, $\Phi_{\mix}^{\hor}(\hat{\bm}_a^0)$ with a second time-step of $0.025\tau$ (while keeping the testing matrix and initial conditions fixed). If the largest overlap component difference is smaller than $10^{-5}$ then we accept the computed endpoint and plug it into the objective function~\eqref{eq:min-squares} for evaluation. Otherwise, if it is larger than $10^{-5}$, we halve the RK4 time-steps once more and reiterate until reaching below the $10^{-5}$ threshold. 

To test the validity of the nonreciprocal model, we perform the same optimization problem on the restricted (symmetric) matrix $\mix_R$ which is constrained such that $\mix_R^{\top}=\mix_R$. In this case, there are six independent entries, whereas for the unrestricted (generally nonreciprocal) mixing matrix $\mix$ we have a total of nine entries to fit. In both fits, we impose the same constraints, $|\mix_{\mu\nu}|\leq 6$, and include no additional penalties. For clarity, the symmetric matrix is optimized independently and is only produced for later comparison and diagnostics. Finally, we run the respective optimization algorithms and retain and record all the values for the solutions with the smallest training objectives.

\paragraph*{Model performance--- }
At each horizon, both fitted models ($\hat{\mix}$ and $\hat{\mix}_R$) are evaluated against the 380 test-ECs from $\test$. Denote the target $\mathbf y_a=\bm_a^{\hor}$ and its prediction by $\widehat{\mathbf y}_a=\Phi_{\hat{\mix}}^{\hor}(\bm_a^0)$. Next, we run the standard statistical scores: 
\begin{align}
\operatorname{RMSE}= \left[\frac{\sum_{a\in\test}{\tilde{g}}_a \lVert\widehat{\mathbf y}_a-\mathbf y_a\rVert^2} {3\sum_{a\in\test}{\tilde{g}}_a} \right]^{1/2}\quad\textrm{and} \quad
R^2= 1- \frac{\sum_{a\in\test}{\tilde{g}}_a \lVert\widehat{\mathbf y}_a-\mathbf y_a\rVert^2} {\sum_{a\in\test}{\tilde{g}}_a \lVert\mathbf y_a-\overline{\mathbf y}_{\tilde{g}}\rVert^2}\quad \textrm{with}\quad \tilde{g}_a=\frac{|\test|}{3\tilde{n}_{\mu(a)}}\,,
\end{align}
where $\tilde{n}_{\mu}$ counts the number of test ECs labeled with phenotype $\mu$. This choice of $\tilde{g}_a$  balances the scores by accounting for the distinct phenotype counts within $\test$ (as done above for the objective function with $\train$); while choosing $\tilde{g}_a=1$ leads to ordinary scores. In the denominator of $R^2$, $\overline{\mathbf y}_{\tilde{g}}:=\sum_{a\in\test}\tilde{g}_a{\mathbf y}_a/(\sum_{a\in\test}\tilde{g}_a)$. This denominator effectively measures how much the targets vary around their phenotypic-population pondered average. Fig.~\ref{fig:A5} reports these statistics, with RMSE computed for each cell using $\hor=\hor^*=5$ hours, and $R^2$ as a function of the horizon values in our designated range. For comparison, in the $R^2$ plot we also compute the $R^2$ for the persistence model, which simply takes the $\bm^{\hor}_{\textrm{pers.}}=\bm^0$, and thus requires no fitting and no DMFT integration.

\begin{figure}
    \centering
    \includegraphics[width=0.85\linewidth]{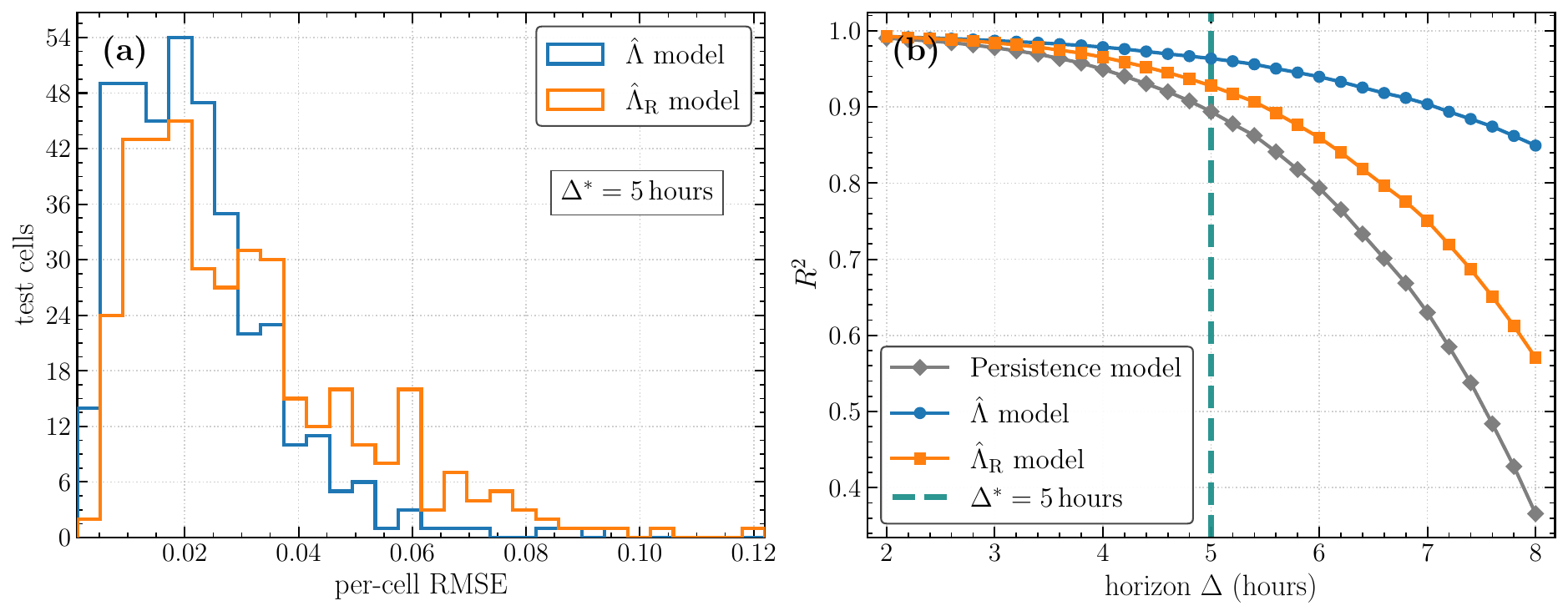}
    \caption{{\textbf{Model performance.}} Panel {\bf (a)} shows the distribution of RMSE values for test ECs under the nonreciprocal model (in blue), which is the goal model of this paper; and under the reciprocally-constrained model (in orange). Both give a range of valid RMSE. Computations are carried out in fixed horizon value $\hor^* = 5$ hours. Panel {\bf (b)} shows the range of $R^2$ values obtained for the unrestricted (nonreciprocal, in blue), restricted (reciprocal, in orange) and persistence (in gray) models across the designated range of horizon values. We note that the nonreciprocal model provides the most accurate predictions in the desired $\hor$ regime.}    \label{fig:A5}
\end{figure}

We note that test ECs are duly excluded from the mixing-matrix estimation (as well as the standardization of expressions and phenotype spin-like memories). However, the upstream RNA-velocity inference uses the full shared cell ensemble. Thus, the reported scores allow us to test our predictions only conditioned on this upstream analysis.

Finally, we note that the nonreciprocal (unrestricted) model contains a total of $p_{\textrm{U}}:=9$ adjustable parameters, while its reciprocal counterpart contains only $p_{\textrm{R}}:=6$ (which follows from the condition that $\mix=\mix^{\top}$). This difference in the number of adjustable parameters could result in a smaller training error in favor of the nonreciprocal model, simply due to the additional parameters at play. To assess this possibility, we perform a final test on the nonreciprocal and reciprocal models based on the $F-$test statistic.

Recall our objective function~\eqref{eq:min-squares} and compute its value against either $\hat{\mix}$ or $\hat{\mix}_R$ for the  nonreciprocal/unrestricted or the reciprocally-restricted models, respectively, for each horizon value $\hor$. Denote by $\textrm{RSS}_{\textrm{U}}(\hor)
=\mathcal{L}_{\hor}(\hat{\mix}(\hor))$ and $\textrm{RSS}
_{\textrm{R}}(\hor)=\mathcal{L}_{\hor}(\hat{\mix}_{\textrm{R}}(\hor))$ the training squared residuals sum computed on each respective model. The $F(\hor)$ statistic is then given by~\cite[Chap.~3]{bates1988nonlinear}
\begin{align}
F(\hor)=\left(\frac{\textrm{RSS}_{\textrm{R}}(\hor)-\textrm{RSS}
_{\textrm{U}}(\hor)}{\textrm{RSS}
_{\textrm{U}}(\hor)}\right)\frac{3\lvert \mathcal{T}\rvert-p_{\textrm{U}}}{p_{\textrm{U}}-p_{\textrm{R}}}\,,\label{eq:F-test}
\end{align}
where the value $3\lvert \mathcal{T}\rvert$ reflects the fact that for each cell in the training set, we obtain three overlap values, which together give rise to the complete sample size. In short, Eq.~\ref{eq:F-test} is effectively a ratio that balances the improvement of the model with additional parameters against the restricted model in a duly variance-normalized way. Moreover, the factor on the right of~\eqref{eq:F-test} rescales the computation according to how many extra parameters are included in the unrestricted versus restricted models. 
Large values of $F$ represent an improvement of the unrestricted/nonreciprocal (${\textrm{U}}$) model relative to the error scale, and which readily accounts for its extra number of parameters.

\begin{figure}
    \centering
    \includegraphics[width=0.9\linewidth]{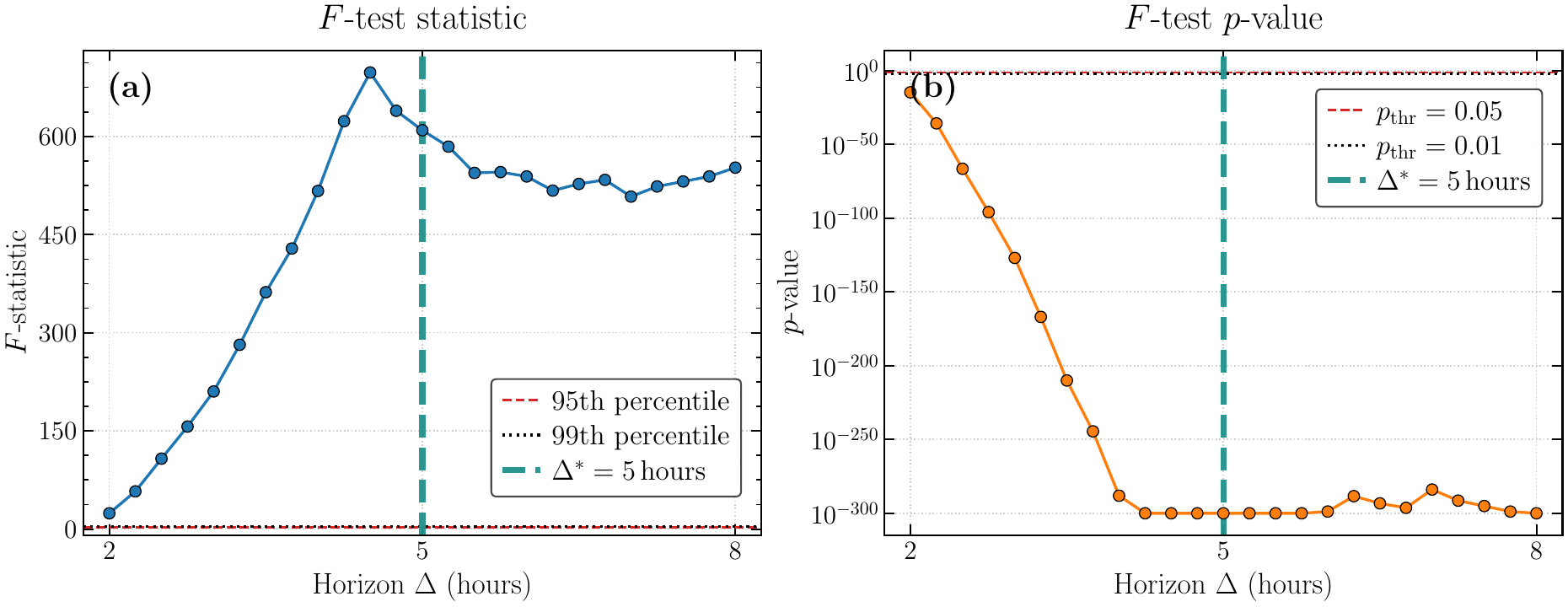}
\caption{{\textbf{$F-$test for the nonreciprocal versus reciprocal model fits.}} \textbf{(a)} Shows the values of $F(\hor)$ comparing the unrestricted (nonreciprocal) model of nine adjustable parameters with the reciprocal (six-parameter) one. Red dashed and black dotted lines display the lower bounds for the 95th and 99th percentiles of the null hypothesis values. \textbf{(b)} Shows the $p-$values associated to the observed $F(\hor)$, with the red dashed and black dotted lines their respective $5\% $ and $1\%$ thresholds. Green vertical dashed line marks the location of the selected $\hor^*=$5 hours.}   \label{fig:A6}
\end{figure}

The null hypothesis in our case is the reciprocal mixing matrix. Under this hypothesis, we compare the resulting $F$ statistic against the Fisher-Snedecor distribution $\mathsf{F}(\nu_1,\nu_2$) with degrees of freedom $\nu_1=p_{\textrm{U}}-p_{\textrm{R}}=3$ and $\nu_2=3\lvert\mathcal{T}\rvert=4539$. In Fig.~\ref{fig:A6}a, we report the values obtained for the $F$ statistic across the explored range for the horizon parameter $\hor$, which are of order $10^2$ across said range. We compare against the theoretical 95th and 99th percentiles, which numerically give lower boundaries of the upper $5\%$ and $1\%$ tails at values
$2.61$ and $3.79$ (shown in red dashed and black dotted lines), respectively. This shows that the $F(\hor)$ obtained departs significantly from the null hypothesis, hence providing statistical evidence for the improvement of the performance of the unrestricted/nonreciprocal model.

We complete this analysis in Fig.~\ref{fig:A6}b by computing the respective $p-$values for all our compared fits, which are directly obtained as $p(\hor)=\textrm{Pr.}\left[X\geq F(\hor)\right]$, with $X\sim \mathsf{F}(3,3\lvert\mathcal{T}\rvert)$. Similarly, this shows that the probability of observing an improvement due only to the additional adjustable parameters and with an $F$ value at least as large as the actual measured one is vanishingly small, with $p\sim 10^{-300}$.

\subsection{Other model outputs}
\label{subsec:model-outputs}

\paragraph*{Correlation (Gram) matrix --- } Before completing the fit for the mixing matrix $\mix$ described above, the empirical Gram matrix $\mathcal Q$ (defined in Eq.~\eqref{eq:Gram-matrix-sm}) is a readily accessible measurable quantity for the ECs in either the $\train$ or the $\test$ (see Fig.~\ref{fig:A7}). Recall that we keep the memories explicitly derived from the $\train$ set at the start of the fitting process and that their respective expression and spin-like vectors remain unchanged across fitting horizons. In short, the empirical memories serve as reference expression patterns, and the stationary state and dynamical properties of the fitted model are determined separately.

\begin{figure}
    \centering
    \includegraphics[width=0.7\linewidth]{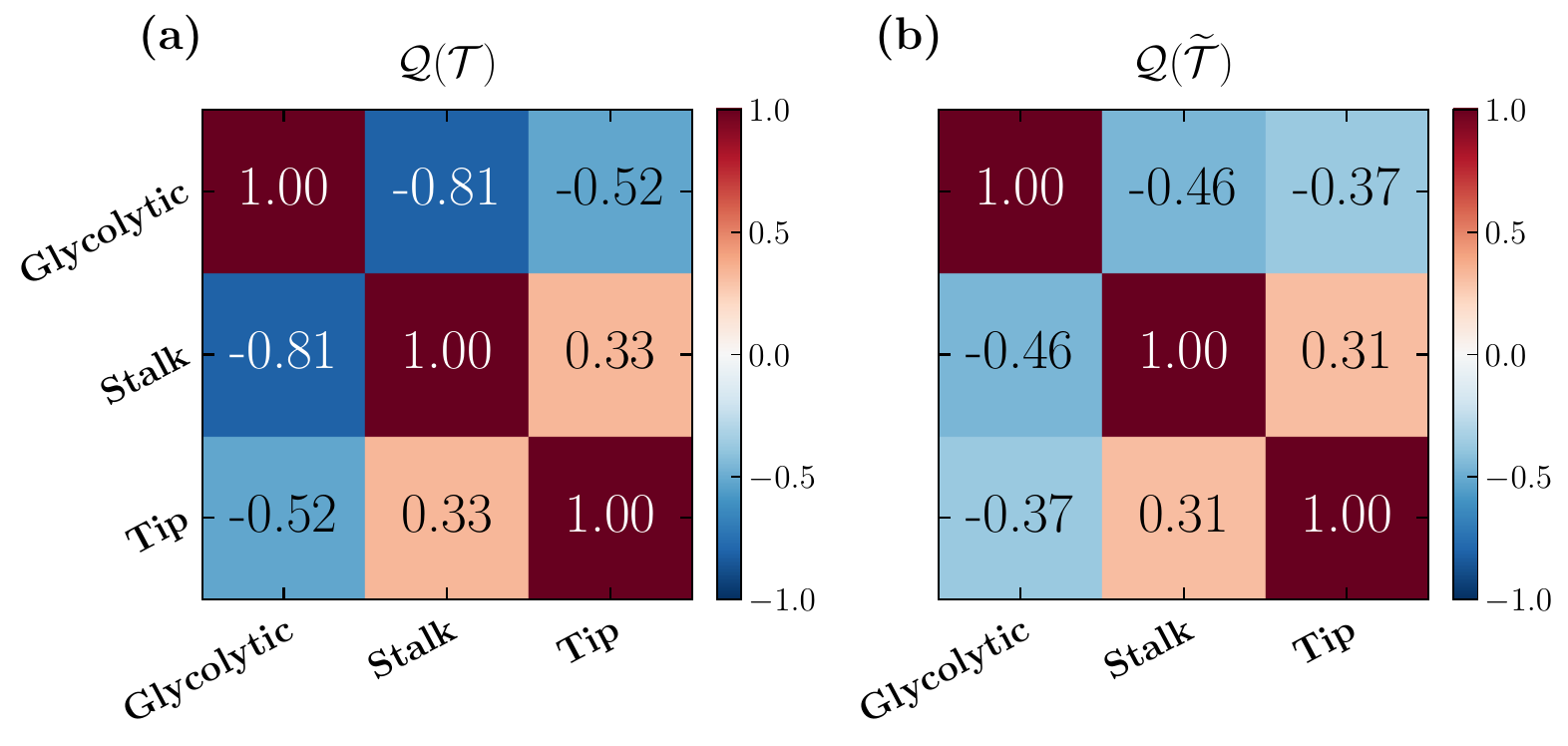}
    \caption{{\textbf{Correlation (Gram) matrices.}} Panels {\bf (a)} and {\bf (b)} respectively show the correlation matrices $\mathcal{Q}(\train)$ and $\mathcal{Q}(\test)$ computed over the training and test partitions of the ECs dataset. In the partition used in our pipeline, we obtained: $(n_{\textrm{G}},n_{\textrm{S}},n_{\textrm{T}}) = (1056,275,185)$  occupancies in $\train$ and $(\tilde{n}_{\textrm{G}},\tilde{n}_{\textrm{S}},\tilde{n}_{\textrm{T}}) = (264,69,47)$ in  $\test$. }   \label{fig:A7}
\end{figure}

\paragraph*{Fixed points and local stability --- }
For each fitted $\hat{\mix}$, we search numerically for stationary solutions satisfying $\bm^*=\boldsymbol{F}(\bm^*)$, where $\boldsymbol{F}(\cdot)$ here denotes the nonlinear contribution of the right hand side of the DMF Eqs.~\eqref{eq:DMFT-m-overlaps-vs-group-fields} (which depends on the choice of $\hat{\mix}$). Here, $\bm^*$ defines a fixed point of the DMFT. Numerical searches for fixed points are carried out for a given $\hat{\mix}$ fit by initializing at the three memory locations, their negatives, and $150$ other randomly generated points in the feasible overlap domain. The trivial fixed point is analyzed separately. Candidate roots are retained when $\lVert\boldsymbol{F}(\bm^*)-\bm^*\lVert^2\leq10^{-8}$, and candidate roots such that $\lVert \bm_1^*-\bm^*_2\rVert<10^{-6}$ are treated as duplicates.

For a given fixed point, its local stability is determined by the properties of the Jacobian,
\begin{align}
\mathcal J_{\hor}(\bm^*)= \frac{1}{\tau} \left[ -I+ \mathfrak H_{3\times4} D(\bm^*) \mathfrak H_{3\times4}^{\intercal} \hat{\mix}(\hor) \right]\,, 
\end{align}
where $D(\bm^*)$ is a $4\times 4$ diagonal matrix with entries $\rho_X\operatorname{sech}^2\left[h_X(\bm^*)\right]$ for the indexes $X\in\{A,B,C,D\}$, which arise from the spin groups introduced in~\ref{subsubsec:empirical-corrs}-\ref{subsubsec:dmft_groups}. We note that the Jacobian depends on the $\hor$, and so does the presence or absence of a given fixed point, $\bm^*$. The Jacobian for the trivial fixed point, which we termed $\mathcal{K}-I$, was already introduced in~\ref{subsubsec:trivial_solution} and is analytically tractable.

A fixed point $\bm^*$ is locally stable (at large times) when every eigenvalue of its associated Jacobian $\mathcal J_{\hor}(\bm^*)$ has a negative real part and unstable when at least one has a positive real part. For each fixed point, the real parts of its eigenvalues are numerically compared with a tolerance of $10^{-7}/\tau$. Value differences within this tolerance range are classified as marginal. Fig.'s~\ref{fig:FIG2}(a-b) in the main text present the spectra at the origin (left) and at the nonzero fixed points (right), using the respective dimensionless eigenvalues $\{\tau\kappa\}$ for each fixed point as found for each $\hat{\mix}(\hor)$, which depends on the chosen horizon $\hor$.

Importantly, imaginary eigenvalues describe oscillatory perturbations whose amplitudes grow or decay according to their real parts. For example, damped oscillations may terminate at a stable fixed point, whereas special cases result in apparent cycles. As discussed in the main text, for the values of $\hor>4$ we report such candidate cycles. In general, the predictions provided by studying the Jacobian's properties are only local, and nonlinearities can give rise to trajectories whose behavior is not fully determined by such properties. In what follows, we describe a numerical approach to identify persistent oscillations in our system.

\paragraph*{Numerical identification of persistent oscillations.}

To test for persistent oscillations, we take every horizon value $\hor$ within our range and every fitted $\mix$, and then we examine ten initializations: the three memory overlap vectors and their negatives, the origin, one specified EC from the test set $\test$, and two additional randomly selected ECs from the training set $\train$. These initial conditions lie within the feasible overlap domain and are held fixed across the chosen $\hor$ values. In our PC-projected maps shown in Fig.~\ref{fig:FIG4} from the main text, we took the initial condition that starts from the aforementioned test EC. 

We initially run trajectories for times up to $500$ in physical units of hours and then extend the integration time for nonstationary trajectories to assess their late-time behavior. Trajectories are integrated using \texttt{solve\_ivp}, with relative and absolute tolerances of $10^{-9}$ and $10^{-11}$, respectively. To characterize the late-time behavior and avoid any potential transients, we focus our analysis on the final quarter of each diagnostic integration. Within this interval, we identify the overlap component with the largest peak-to-peak range and store its successive maxima. Their mean spacing defines our estimated periods,
\begin{align}
T(\hor)=\frac{1}{n_p-1}
\sum_{j=1}^{n_p-1}(t_{j+1}-t_j),
\end{align}
where $n_p$ is the number of detected peaks, and $t_j$ is the physical time of the $j$th peak. We identify persistent oscillations by requiring several clear cycles in the late-time trajectory, with approximately constant amplitude and period. We also check that their amplitude exceeds small numerical fluctuations and that the temporal sampling adequately resolves each cycle. These checks help distinguish sustained oscillations from decaying transients or numerical artifacts.

A trajectory is classified as stationary when its largest late-time component range is below $10^{-4}$ and its final dimensionless speed satisfies $\lVert\boldsymbol{F}_{\hor}(\bm)\rVert<10^{-6}$. If a trajectory does not satisfy either criterion, then it is deemed {\em unresolved}. This includes cases with too few observed peaks, insufficient temporal resolution, or amplitudes and peak spacings that have not settled. Nonstationary trajectories are checked over longer observation intervals, up to a maximum duration of $10^4\tau$.  The plots in Fig.~\ref{fig:A8} show only the first $500\, \mathrm{hours}$.

Fig.~\ref{fig:FIG2}c from the main text shows the periods predicted by the unrestricted (nonreciprocal) model across the chosen horizon values, using the same initial condition throughout. Trajectories ending at a fixed point are represented at zero as a graphical convention, and no unresolved cases are found within this range.  
Since each horizon corresponds to a separately inferred matrix $\hat{\mix}(\hor)$, this comparison informs us about the sensitivity of the model's oscillatory behavior to the horizon values $\hor$.

\begin{figure}
    \centering
    \includegraphics[width=0.95\linewidth]{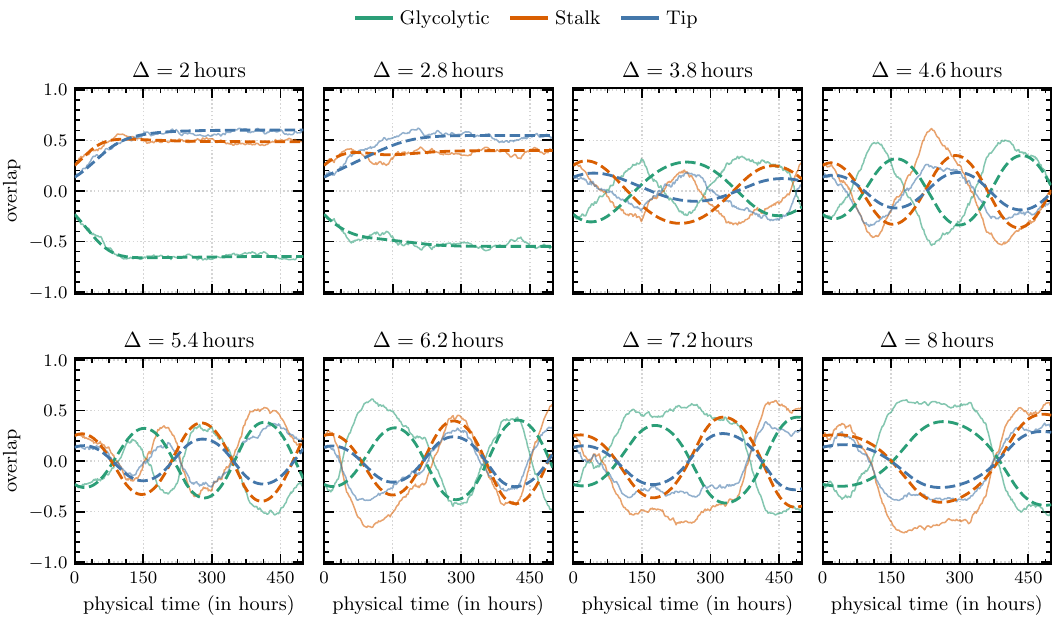}
    \caption{\textbf{Deterministic and stochastic runs.}
    Phenotypic-overlap trajectories generated by the our nonreciprocal model and fitted at the indicated horizon values, $\hor$, in units of hours. Thick dashed curves show the DMFT predictions, while thin solid curves show the corresponding Glauber realizations. Colors encode the overlaps with the three relevant phenotype memories: Glycolytic, Stalk, and Tip.}
    \label{fig:A8}
\end{figure}

\fi

\vspace{1em}
\twocolumngrid


\end{document}